\documentclass[traditabstract]{aa}

\usepackage{natbib}
\usepackage{graphicx}
\usepackage{commath}
\usepackage{siunitx}

\usepackage{txfonts}

\newcommand{\uHz}[1]{\SI{#1}{\micro\hertz}}

\newcommand{\be}{\begin{equation}}
\newcommand{\ee}{\end{equation}}

\def\ba{\begin{eqnarray}}
\def\ea{\end{eqnarray}}

\def\msol{M_\odot}

\def\he3{^3He}

\def\ltsima{$\; \buildrel < \over \sim \;$}
\def\simlt{\lower.5ex\hbox{\ltsima}}
\def\gtsima{$\; \buildrel > \over \sim \;$}
\def\simgt{\lower.5ex\hbox{\gtsima}}
\def\vel{{\rm v}}
\def\velvec{{\rm \mathbf v}}

\def\Fsingle{F_{\rm wave}}

\newcommand{\cp}{\citep}
\newcommand{\ct}{\citet}

 \renewcommand{\vec}[1]{\mathbf{#1}}

\newcommand{\AvgS}[1]{\left< {#1} \right>_{\mathcal{S}}}
\newcommand{\AvgST}[1]{\left< {#1} \right>_{\mathcal{S},t}}
\newcommand{\AvgT}[1]{\left< {#1} \right>_t}

\newcommand{\Vrms}{\vel_\mathrm{rms}}
\newcommand{\domdkh}{\delta \omega \, \delta k_{\rm h}}

\usepackage{xcolor}
\definecolor{xlinkcolor}{cmyk}{1,1,0,0}
\usepackage[
     colorlinks=true,    % false: boxed links; true: colored links
     linkcolor=xlinkcolor,     % color of internal links
     citecolor=xlinkcolor,     % color of links to bibliography
     filecolor=xlinkcolor,  % color of file links
     urlcolor=xlinkcolor      % color of external links
]{hyperref}
\RequirePackage[normalem]{ulem} %DIF PREAMBLE
\RequirePackage{color}\definecolor{RED}{rgb}{1,0,0}\definecolor{BLUE}{rgb}{0,0,1} %DIF PREAMBLE
\begin{document}

\title{Two-dimensional simulations of solar-like models with artificially enhanced luminosity. II. Impact on internal gravity waves}
\titlerunning{Two-dimensional simulations of solar-like models with artificially enhanced luminosity}

\author{A.~Le~Saux \inst{1,2}, T.~Guillet\inst{1}, I.~Baraffe \inst{1,2}, D.~G.~Vlaykov\inst{1}, T.~Constantino \inst{1}, J.~Pratt\inst{3}, T.~Goffrey\inst{4}, M.~Sylvain\inst{1}, V.~Réville\inst{5}, A.~S.~Brun\inst{6}}
\authorrunning{Le Saux et al.}

\offprints{A. Le Saux}

\institute{
University of Exeter, Physics and Astronomy, EX4 4QL Exeter, UK
(\email{al598@exeter.ac.uk})
\and
\'Ecole Normale Sup\'erieure, Lyon, CRAL (UMR CNRS 5574), Universit\'e de Lyon, France
\and
Department of Physics and Astronomy, Georgia State University, Atlanta GA 30303, USA
\and
Centre for Fusion, Space and Astrophysics, Department of Physics, University of Warwick, Coventry, CV4 7AL, UK
\and
Institut de Recherche en Astrophysique et Planétologie, CNRS, UPS, CNES, Toulouse, France
\and
AIM, CEA, CNRS, Universités Paris et Paris-Saclay, 91191 Gif-sur-Yvette, Cedex, France
}

\date{}

\abstract{Artificially increasing the luminosity and the thermal diffusivity of a model is a common tactic adopted in hydrodynamical simulations of stellar convection. In this work, we analyse the impact of these artificial modifications on the physical properties of stellar interiors and specifically on internal gravity waves. We perform two-dimensional simulations of solar-like stars with the MUSIC code. We compare three models with different luminosity enhancement factors to a reference model. The results confirm that properties of the waves are impacted by the artificial enhancement of the luminosity and thermal diffusivity. We find that an increase in the stellar luminosity yields a decrease in the bulk convective turnover timescale and an increase in the characteristic frequency of excitation of the internal waves. We also show that a higher energy input in a model, corresponding to a larger luminosity, results in higher energy in high frequency waves. Across our tests with the luminosity and thermal diffusivity enhanced together by up to a factor of $10^4$, our results are consistent with theoretical predictions of radiative damping. Increasing the luminosity also has an impact on the amplitude of oscillatory motions across the convective boundary. One must use caution when interpreting studies of internal gravity waves based on hydrodynamical simulations with artificially enhanced luminosity.}


\maketitle
%
%%%%%%%%%%%%%%%%%%%%%
%   SECTION 1
%%%%%%%%%%%%%%%%%%%%%
%
\section{Introduction}
Internal gravity waves (IGWs) have an impact on the internal structure and evolution of stars as they can  transport angular momentum, energy, and chemical elements in stably stratified media. They are well studied in geophysical \citep[see for example][]{Fritts2003, Sutherland2010} and experimental contexts \citep[see for example][]{Plumb1978, Leard2020}, but their properties in stellar interiors are still poorly constrained. Their detection and characterisation in stars, particularly in solar-like stars, is very difficult and remains a long-standing challenge \citep{Garcia2007, Appourchaux2010}. As IGWs propagate in the radiative core of the Sun, their detection could provide key information about the deep solar interior.
They are suggested as a possible mechanism to explain the uniform rotation of the solar radiative zone \citep{Kumar1999, Charbonnel2005} and  to explain the observed differential rotation in evolved stars \cp[see for example][]{Fuller2014}.

Hydrodynamical simulations provide an alternative `laboratory' for studying IGWs in stellar interiors and testing theoretical predictions. However, the wide range of time and length scales that characterise stellar interiors is a challenge for numerical simulations. In the case of the Sun, timescales range from several minutes, for the dynamical timescale, to several million years, for the thermal timescale. In addition, the study of IGWs requires simulating a large region of a star that includes the convective and the radiative zones because these waves are produced by the non-linear interactions between these two regions. The two main mechanisms responsible for IGW generation are the Reynolds stress resulting from turbulent convection \citep{Stein1967, Press1981, Goldreich1990, Lecoanet13} and the convective penetration in the underlying stably stratified region \citep{Rieutord1995, Montalban2000, Pincon2016}. To overcome numerical difficulties inherent to stellar hydrodynamics, such as numerical stability and thermal relaxation, an artificial increase in the stellar luminosity and in the thermal diffusivity by several orders of magnitude is a commonly used tactic \citep{Meakin2007, Brun2011, Rogers2013, Jones2017, Edelmann2019, Horst2020}. Indeed, these modifications can increase the numerical stability of the simulation and accelerate the thermal relaxation.
In a first study \citep[][hereafter Paper I]{Baraffe2021}, we performed two-dimensional fully compressible time-implicit simulations of convection in a solar-like model using the MUSIC code. We analysed the impact  of increasing the stellar luminosity and thermal diffusivity  on convective penetration (or overshooting) at the base of the convective envelope of a solar-like model. We showed that these modifications can significantly affect the structure of the overshooting layer. However, the impact on the properties of IGWs had never been studied.
In a recent work, \citet{Lecoanet2019} suggested that artificially increasing the luminosity in numerical simulations makes it difficult to study wave excitation and propagation since it may change the wave physics. This second study is thus specifically devoted to estimating the impact of the luminosity enhancement on waves. Our goal is to  answer the question of whether numerical simulations using such artificial modifications can be used to predict the properties of waves in more realistic systems. \\

We start in Sect. \ref{sec:num_sim} by describing the simulations used for this study. The effects of the boost on fluid velocities are presented in Sect. \ref{sec:velocities}. In Sect. \ref{sec:CZ} we study the properties of IGWs and in Sect. \ref{sec:IGW} their excitation and damping. Finally, in Sect. \ref{sec:theory} we compare our results with theoretical predictions.

%%%%%%%%%%%%%%%%%%%%%
%   SECTION 2
%%%%%%%%%%%%%%%%%%%%%

\section{Numerical simulations}
\label{sec:num_sim}

\subsection{Numerical models}
We performed two-dimensional simulations of convection in a solar-like model  with MUSIC \citep{Viallet11, Viallet13, Viallet16, Geroux16, Goffrey17, Pratt2016, Pratt2017, Pratt2020}.
The numerical simulations are described in detail in Paper I. We summarise below the main information relevant to the present analysis.
The code solves the hydrodynamical equations in a fully compressible medium

\begin{equation}
\frac{\partial \rho}{\partial t} = - \vec{\nabla} \cdot (\rho \vec \velvec),
\end{equation}

\begin{equation}
\frac{\partial \rho e}{\partial t} = - \vec{\nabla} \cdot (\rho e \vec\velvec) - p \vec{\nabla} \cdot \vec \velvec - \vec{\nabla} \cdot \vec{F_r},
\end{equation}

\begin{equation}
\frac{\partial\rho \vec\velvec}{\partial t} = - \nabla \cdot (\rho \vec \velvec \otimes \vec \velvec) - \vec{\nabla}p + \rho\vec{g},
\end{equation}
where $\rho$ is the density, $e$ the specific internal energy, $\vec \velvec$ the velocity field, $p$ the gas pressure, and $\vec{g}$ the gravitational acceleration. For the stellar simulations considered in this work, the major heat transport that contributes to thermal conductivity  is radiative transfer characterised by the radiative flux  $\vec{F_r}$, given within the diffusion approximation by

\begin{equation}
\vec{F_r} = - \frac{4acT^3}{3\kappa \rho} \vec{\nabla} T = - \chi \vec{\nabla} T,
\label{eq:radiative_flux}
\end{equation}
with $\kappa$ is the Rosseland mean opacity of the gas and   $\chi$  the radiative conductivity.
The opacities are interpolated from the OPAL tables \citep{Iglesias1996} for solar metallicity and the equation of state is based on the OPAL tables of \citep{Rogers2002}, which are appropriate for the description of solar-like interior structures.
The numerical domain is a spherical shell assuming azimuthal symmetry and covered by spherical coordinates: the radius $r$, and the co-latitude $\theta$. In multi-dimensional simulations of stellar interiors it is usual to refer to a direction perpendicular to the radial one as the horizontal direction. In the following, we identify the horizontal direction as corresponding to the co-latitude ($\theta$). The radial domain extends from $r = 0.4 R_{\rm tot}$ to $r = 0.9 R_{\rm tot}$, this corresponds to the region between the two hatched regions on Fig. \ref{fig:BV}. The domain in the co-latitudinal direction ranges from $0$ to $\pi$, including the full hemisphere. We use a uniform grid resolution of $r\times \theta = 512 \times 512$ cells. This provides a good resolution of the pressure scale height at the Schwarzschild boundary $H_{P,CB}/\Delta r \sim 92$, with $\Delta r = 5.5 \times 10^7$ cm (550 km) the radial grid spacing.
In terms of boundary conditions, we impose a constant radial derivative on the density on the inner and outer radial boundaries, as discussed in \citet{Pratt2016}. For the reference model \textit{ref}, the energy flux at the inner and outer radial boundaries are set to the value of the energy flux at that radius in the one-dimensional stellar evolution model used as initial model (see Paper I for details). For the artificially boosted simulations, the energy flux, and equivalently the luminosity, at the boundaries is multiplied by an enhancement factor, and the Rosseland mean opacities $\kappa$ in MUSIC are decreased by the same factor. In this work we analyse the impact of enhancing the luminosity and thermal diffusivity by factors of $10$, $10^2$, and $10^4$.
In velocity, we impose on the radial boundaries non-penetrative condition for the radial velocity and stress-free boundary condition for the angular velocity. At the boundaries in $\theta$, because of the extension of the angular domain to the `poles', we use reflective boundary conditions for the density and energy, meaning that they are mirrored at the boundary. We adopted a stress free boundary condition for the radial velocity and a reflecting boundary condition for the angular velocity to ensure it is equal to zero at the boundary. \\

\begin{table}[t]
   \caption{Properties of the initial reference model.}
   \label{tab0}
   \centering
   \begin{tabular}{c c c c c }
     \hline \hline
     $M/\msol$ &  $L_{\rm star}/L_\odot$ & $R_{\rm tot}$ (cm) &  $r_{\rm conv}/R_{\rm tot}$ & $H_{P,{\rm conv}}$ (cm) \\
      \hline
      1 &  1.07 & 5.64491 $\times 10^{10}$ & 0.6734 & 5.086 $\times 10^{9}$ \\
      \hline
   \end{tabular}
    \tablefoot{
    With mass $M$, luminosity $L_{\rm star}$, radius $R_{\rm tot}$, depth of the convective envelope $r_{\rm conv}$ and pressure scale height at the convective boundary $H_{P,{\rm conv}}$.}
\end{table}

The initial stellar model for the two-dimensional simulations is that of a solar mass star with a radius and luminosity close to the solar values. It is constructed with a one-dimensional stellar evolution code in such a way that it allows an artificial increase in the luminosity and of the thermal diffusivity (see details in Paper I).
The properties of the initial reference model are summarised in Table \ref{tab0} and the characteristics of the four numerical models used in this work are presented in Table \ref{tab1}. The convective turnover time $\tau_{\rm conv}$ is given by
\begin{equation}
\tau_{\rm conv} =  \left< \int_{r_{\rm conv}}^{r_{\rm out}} \frac{\dif r} { \Vrms(r,t)  } \right>_t   = \omega_{\rm conv}^{-1},
\label{tau}
\end{equation}
where $r_{\rm out} = 0.9 R_{\rm tot}$ is the outer boundary of the two-dimensional simulations and $r_{\rm conv}= 0.6734 R_{\rm tot}$ is the convective boundary of the stellar model as defined by the Schwarzschild criterion. The brackets $\AvgT{.}$ denotes a time average and it is defined as
\begin{equation}
    \left< f(t) \right>_t \coloneqq \frac{1}{T} \int_0^T f(t) \dif t,
    \label{eq:time_av}
\end{equation}
with $T$ the time of integration. The lower limit of the integral $t=0$ corresponds to the time from which convection is in steady state. We also define the convective turnover frequency $\omega_{\rm conv}$ by the frequency associated with the characteristic timescale $\tau_{\rm conv}$.

\begin{table}[h!]
   \caption{Summary of the two-dimensional simulations.}
   \label{tab1}
   \centering
   \begin{tabular}{l c c c c}
     \hline \hline
     Simulation &  $L/L_{\rm star}$ & $\tau_{\rm conv}^{(a)}$ (s) &  $N_{\rm conv}^{(b)}$ &  $\omega_{\rm conv}^{(c)}$ (\uHz{})  \\
      \hline
      ref &  1 & 8 $\times 10^5$ & 565 & 1.25 \\
       boost1d1 &  10$^1$ & 3.6 $\times 10^5$ & 375 & 2.78 \\
        boost1d2 &  10$^2$ &   1.7 $\times 10^5$ & 450 & 5.88\\
        boost1d4 &  10$^4$ & 3.5 $\times 10^4$& 530 & 28.57\\
      \hline
   \end{tabular}
 \tablefoot{
\tablefoottext{a}{Convective turnover time. See Eq. \eqref{tau} for its definition.}
 \tablefoottext{b}{Number of convective turnover times covered by the simulation once a steady-state convection is reached.}
  \tablefoottext{c}{Convective turnover frequency, defined as the inverse of the convective turnover time, $1/\tau_{\rm conv}$.}
    }
\end{table}

\subsection{Stratification}
\label{sec:stratification}
\begin{figure}[!h]
    \centering
    \includegraphics[width=0.4\textwidth]{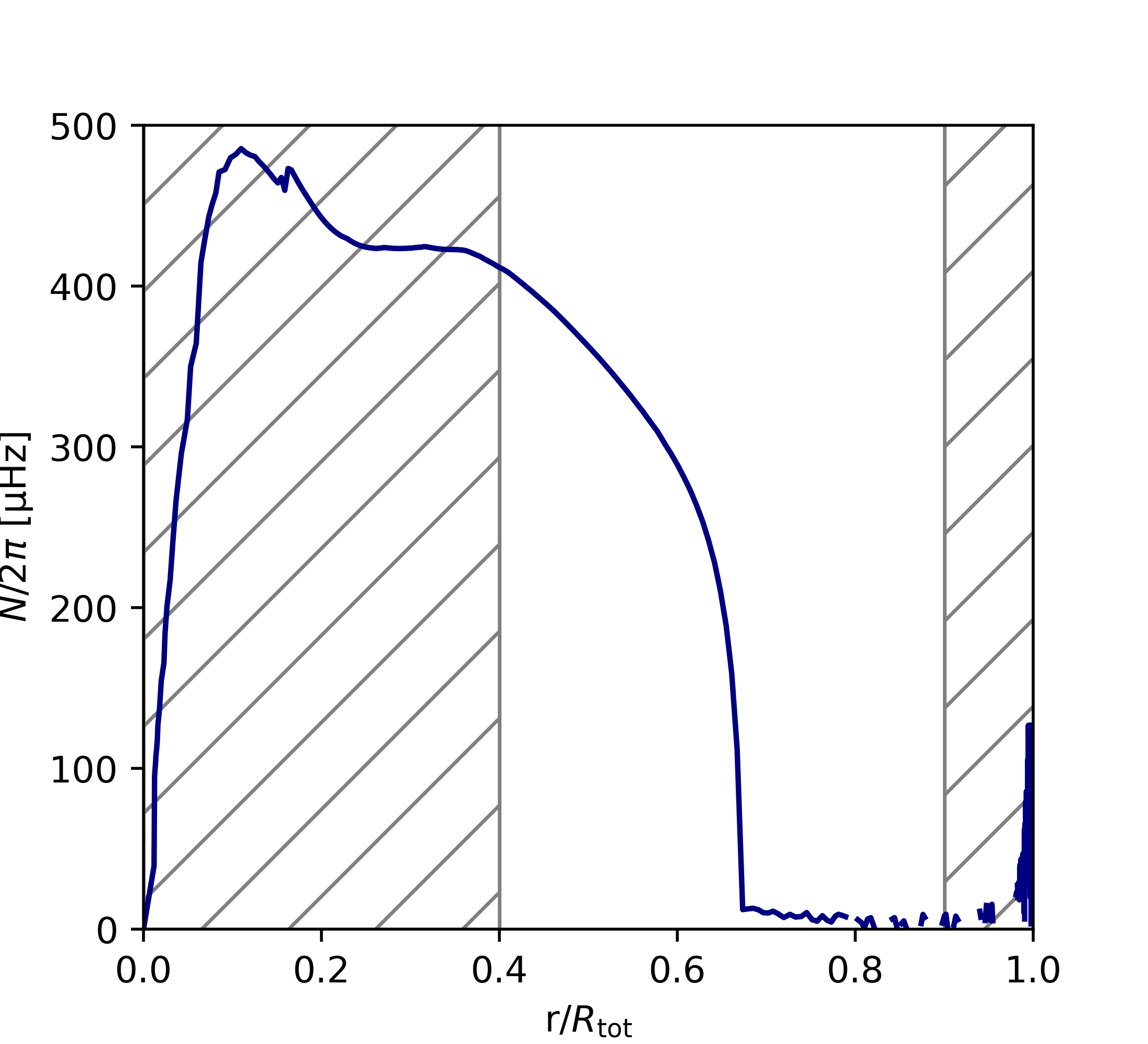}
    \caption{Radial profile of the Brunt-Väisälä frequency for the initial model (see Table~\ref{tab0}). The hatched regions are not considered in the two-dimensional simulations.}
    \label{fig:BV}
\end{figure}

The stratification of a stellar simulation is a key property for the analysis of internal waves and can be characterised by the Brunt-Väisälä frequency, $N$, defined as
\begin{equation}
    N = \sqrt{g \left( \frac{1}{\Gamma_1} \dod{\ln p}{r} - \dod{\ln \rho}{r} \right) },
    \label{eq:BV_freq}
\end{equation}{}
where $\Gamma_1$ is the first adiabatic exponent,
\begin{equation}
  \Gamma_1 = \left(\dpd{\ln \rho}{\ln p}\right)_{\mathrm{ad}}.
\end{equation}{}
This frequency characterises the maximum frequency for vertical oscillations under gravity of a fluid parcel around its equilibrium position \citep{Lighthill78}. Figure \ref{fig:BV} shows the radial profile of the Brunt-Väisälä frequency for the initial model. On this plot, Eq. \eqref{eq:BV_freq} is divided by $2\pi$ in order to get the Brunt-Väisälä frequency in Hz, and not in $\mathrm{rad}.s^{-1}$. Moreover in all this work $\omega$ is a frequency and is expressed in Hz.
As mentioned, the convective boundary of the stellar model is located at $r= 0.6734 R_{\rm tot}$. At a given radius, $r_0$, IGWs with frequencies $\omega \leq N(r=r_0)$ can propagate in the stably stratified region below the convective boundary.

The radial profile of $N$ close to the convective boundary plays an important role for the excitation of IGWs, as shown by the analytical work of \citet{Lecoanet13} who found that the IGW flux at the interface can vary by orders of magnitude depending on the steepness of the Brunt-Väisälä profile at the boundary. This effect was also suggested by \citet{Rogers2015} and confirmed by the numerical simulations of \citet{Couston2017} showing  that  increasing the stiffness of the interface (i.e. stronger stratification) decreases the IGW flux.

%%%%%%%%%%%%%%%%%%%%%
%   SECTION 3
%%%%%%%%%%%%%%%%%%%%%

\section{Velocities}
\label{sec:velocities}

\subsection{Root mean square velocity}
\label{sec:rms_vel}
One of the main effects of artificially enhancing the luminosity of a numerical model is to increase the fluid velocities.
This is readily seen by looking at the root mean square (rms) velocities,
which we compute in this work as the mass-weighted squared velocity:
\begin{equation}
    \Vrms(r) \coloneqq \sqrt{\frac{\AvgST{ \rho \velvec^2}}{\AvgST{\rho}}},
    \label{eq:vrms}
\end{equation}
where $\AvgT{.}$ and $\AvgS{.}$ are the time and angular averages. The temporal average was introduced in Eq. \eqref{eq:time_av}, and the angular average is defined as
\begin{equation}
    \left< g(\theta) \right>_{\mathcal{S}} \coloneqq \frac{1}{4\pi} \int_\mathcal{S} g(\theta) \, 2\pi \sin \theta \dif \theta.
    \label{eq:angular_av}
\end{equation}
We are using  Eq.~\eqref{eq:vrms} as the definition for the $\Vrms$ as it is the most relevant for comparison with analytical models that define the $\Vrms$ and the kinetic energy density based on mass-weighted quantities \citep[see for example Eq.~(47) in][]{Goldreich1977}.

\begin{figure}[!h]
    \centering
    \includegraphics[width=0.5\textwidth]{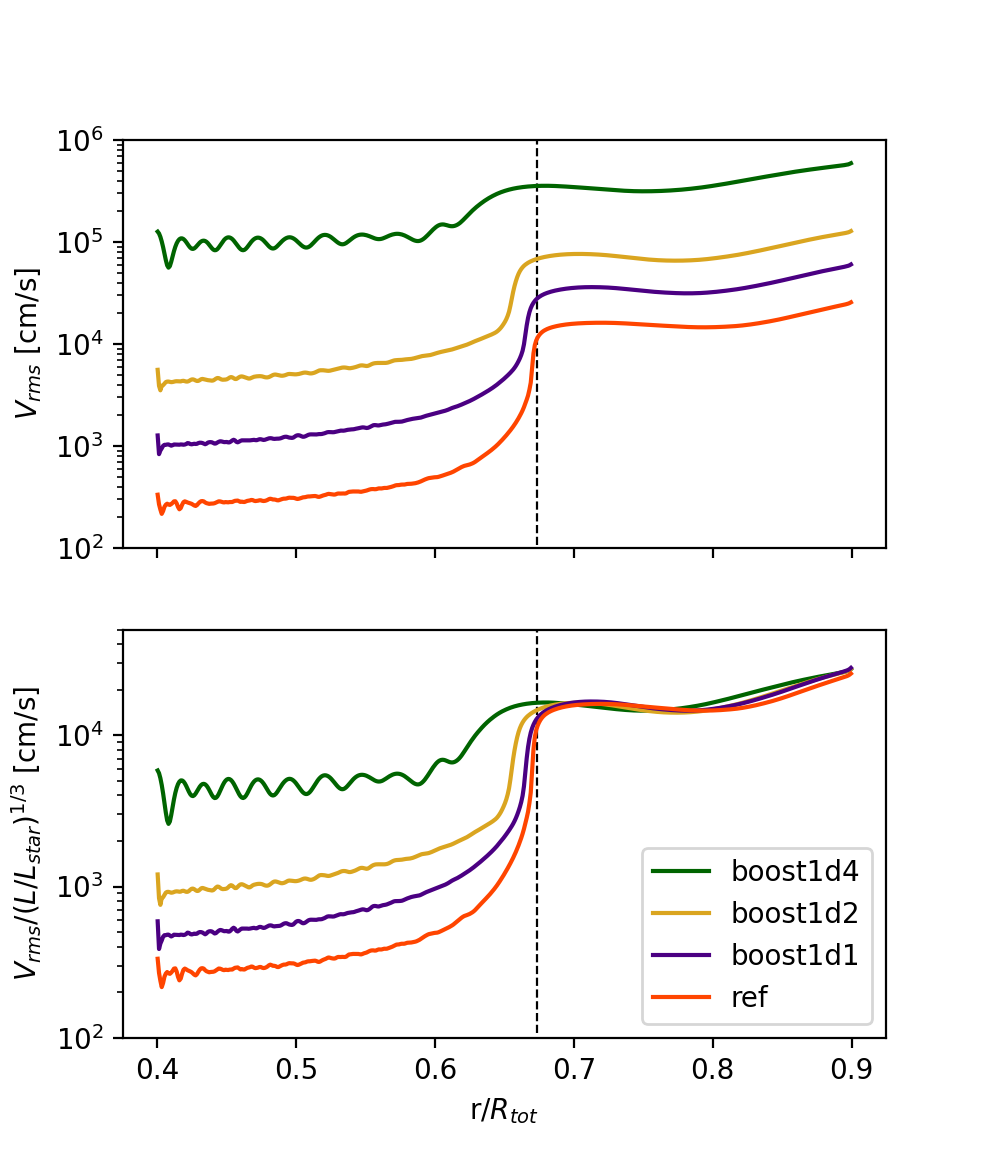}
    \caption{Radial evolution of the rms velocity for the four simulations \textit{ref}, \textit{boost1d1}, \textit{boost1d2}, and \textit{boost1d4}. The convective boundary corresponding to the Schwarzschild boundary from the one-dimensional initial model is indicated by the vertical dashed line.}
    \label{fig:vrms}
\end{figure}

\begin{figure*}[!h]
    \centering
    \includegraphics[width=1.0\textwidth]{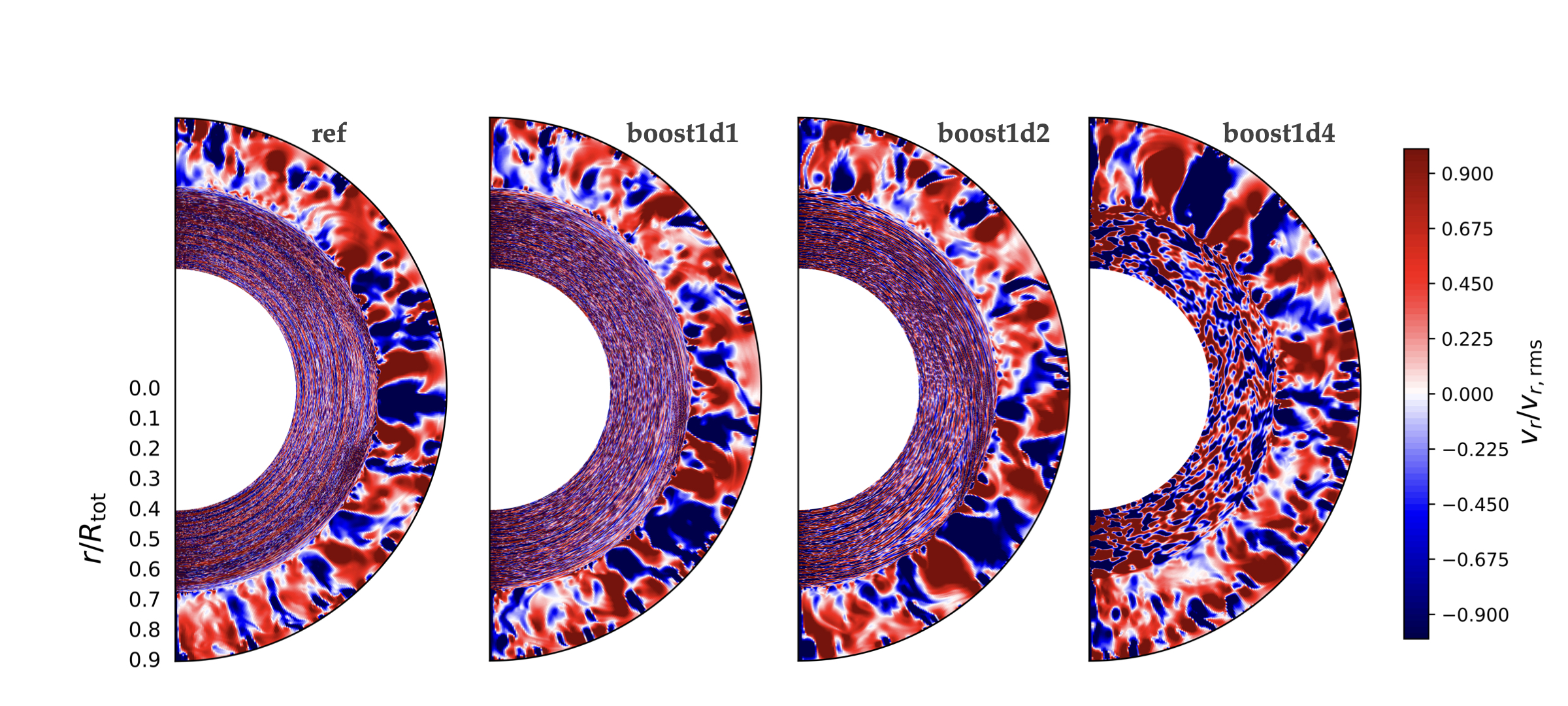}
    \caption{Visualisation of the radial velocity for the four stellar simulations \textit{ref}, \textit{boost1d1}, \textit{boost1d2,} and \textit{boost1d4} as a function of radius, $r$, and co-latitude, $\theta$. The radial velocity is normalised by the rms radial velocity.}
    \label{fig:Ur_2d}
\end{figure*}

Figure \ref{fig:vrms} shows the radial profile of the rms velocity (top panel)  and of the rms velocity normalised by the luminosity enhancement factor to a power $1/3$ (bottom panel). For each simulation the profile is computed over $\sim 375 \tau_{\rm conv}$ and the corresponding value of $\tau_{\rm conv}$ is given in Table \ref{tab1}.
As already discussed in Paper I and shown in Fig. \ref{fig:vrms}, our numerical simulations reproduce the expected scaling of $\Vrms$ in the convective zone with the luminosity
\begin{equation}
    \Vrms \propto L^{1/3}.
    \label{eq:rms-MLT}
\end{equation}
The rms velocities observed in the radiative zone are mainly due to the propagation of internal waves. Figure \ref{fig:vrms} shows that the rms velocity amplitude of the oscillatory motions increases with the luminosity enhancement factor.

Figure \ref{fig:vrms} also shows deeper penetration of the convective motions below the convective boundary and larger overshooting depth with increasing luminosity. These effects were analysed in detail in Paper I.
The extension of the overshooting length with increasing  luminosity is expected from theory \citep{Zahn1991, Rempel2004} and has also been reported in previous numerical simulations \citep[see for example][]{Hotta2017, Kapyla2019}.
Larger velocities in the convective zone and in the overshooting layer with enhanced luminosity factor will impact IGWs  as they are excited by turbulent convection in the convective zone and by penetrating flows across the convective boundary. This will be studied in Sect. \ref{sec:IGW}.

\subsection{Radial velocity}
The excitation of waves at the convective boundary and their propagation in the stably stratified region is well illustrated in Fig. \ref{fig:Ur_2d}, which displays the radial velocity in the two-dimensional plane for the four simulations.
The radial velocity $\mathrm{v}_r$ is normalised by the rms radial velocity, $\vel_{r, \rm rms}(r)$, for better visualisation, as  the amplitude of the velocity in the radiative zone can be several orders of magnitude smaller than its typical value in the convective zone (see Fig. \ref{fig:vrms}).

In the convective envelope, we can clearly see upward (red) and downward (blue) flows characterising the convective motions and these patterns are quite similar in the four simulations. Whereas in the radiative region ($r<0.6734 R_{\rm tot}$) the patterns are different with different luminosity enhancement factors. In \textit{ref} the thin concentric circles are characteristics of IGW wavefronts.
These circular wavefronts are in fact spirals and their inclination with respect to the convective boundary reflects the frequency of the wave \citep{Stein1967}. Indeed, the propagation direction of IGWs is set by their dispersion relation \citep{Vallis2017}:
\begin{equation}
    \frac{\omega^2}{N^2} = \frac{k_{\rm h}^2}{k^2} \coloneqq \cos^2\alpha,
    \label{eq:dispersion}
\end{equation}
where $\omega$ is the wave frequency and $k$ the total wavenumber defined by $k = \sqrt{k_r^2 + k_{\rm h}^2}$, with $k_r$ its radial part and $k_{\rm h}$ its horizontal part defined at a given radius $r$ by:
\begin{equation}
    k_{\rm h}^2 \coloneqq \frac{\ell(\ell+1)}{r^2},
    \label{eq:kh}
\end{equation}
with $\ell \geq 0$ the spherical harmonic degree.
Equation \eqref{eq:dispersion} introduces an angle $\alpha$ defined as the angle between the total and the horizontal wavenumbers. A specific  angle $\alpha$ corresponds to a specific frequency $\omega$. The higher the frequency, the smaller the angle $\alpha$ and the less horizontal the wavefront. This is known as the St. Andrews cross \citep{Sutherland2010}. This dispersion relation also implies the condition of IGW propagation, namely $\omega \leq N$.

For \textit{boost1d1} and \textit{boost1d2}, it can be seen in Fig.~\ref{fig:Ur_2d} that the inclination of the wavefronts increases compared to \textit{ref}. This suggests that the dominant waves have higher frequencies. We confirm this pattern in Sect. \ref{sec:IGW}. For the most boosted simulation \textit{boost1d4},  the characteristic spiral structures in the radiative zone are not visible anymore. This will also be discussed in Sect. \ref{sec:IGW}.

%%%%%%%%%%%%%%%%%%%%%
%   SECTION 4
%%%%%%%%%%%%%%%%%%%%%

\section{Radial kinetic energy spectra}
\label{sec:CZ}

\begin{figure*}[!h]
\centering
   \includegraphics[width=1.0\textwidth]{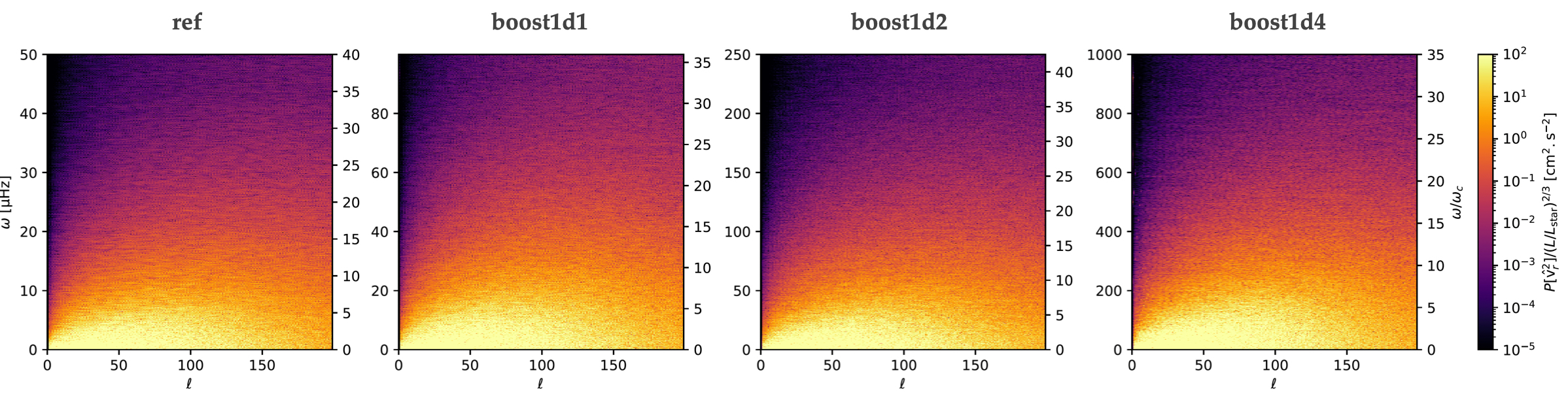}
     \caption{Power spectra of the radial velocity for the four stellar simulations \textit{ref}, \textit{boost1d1}, \textit{boost1d2,} and \textit{boost1d4,} respectively, in the convective zone at depth $r$ = 0.762$R_{\rm tot}$. The spectra were obtained via mode projection on the spherical harmonics basis and a temporal Fourier transform of the radial velocity. The radial velocities are normalised by $(L/L_{\rm star})^{1/3}$  (see text Sect. \ref{conv}).
     }
     \label{Spectrum_Rcz}
\end{figure*}

Given that IGWs result from the dynamical interaction between the convective and radiative zones, the radial kinetic energy spectra in both regions are useful diagnostics to analyse the properties of waves. Since waves are periodic phenomena in space and time,  our analysis is performed in the spectral domain, both spatially and temporally. In the following we use a temporal Fourier transform of the radial velocity $\vel_r(r,\theta, t)$ to obtain a dependence in frequency $\omega$, $\Tilde{\vel}_r(r,\theta, \omega)$. We then perform a projection on the spherical harmonics basis to obtain $\hat{\vel}_r(r,\ell, \omega)$, where $\ell$ is the spherical harmonic degree. The definition we are using for spherical harmonics and Fourier transform are given in Appendix \ref{apdx:sh-ft}.

\subsection{Convective zone}
\label{conv}
We first analyse the power spectrum of the radial velocity in the convective zone for all simulations. An important quantity in this context is the convective turnover timescale, $\tau_{\rm conv}$ (see  Eq. \eqref{tau}), which defines a characteristic frequency $\omega_{\rm conv}$. Equations  \eqref{tau} and \eqref{eq:rms-MLT} imply the scaling relation $\omega_{\rm conv} \propto L^{1/3}$. The values of $\omega_{\rm conv}$ for each simulation are provided in Table \ref{tab1}. As we will see in the following, the characteristic convective turnover frequency is particularly relevant for present analysis.

Figure \ref{Spectrum_Rcz} shows the power spectrum of the radial velocity $P[\hat{\vel}_r^2]$ (see Eq. \eqref{eq:def_power_spectrum} for its definition), at a radius $r = 0.762R_{\rm tot}$, which is located in the bulk of the convective envelope for the four numerical models.  We note that the power spectrum barely depends on the location $r$ within the convection zone, as long as $r$ is far enough from the top and bottom boundaries. Based on the scaling relations for the velocities and the convective frequency with the luminosity enhancement factor, the power spectra displayed in Fig. \ref{Spectrum_Rcz} are calculated with the velocity divided by $(L/L_{\rm star})^{1/3}$. For all simulations,  the power spectrum values range between $10^{-5}$ and $10^2$ $\rm cm^2.s^{-2}$. By also dividing the frequency $\omega$ by  $\omega_{\rm conv}$, providing the same range of normalised frequency for all four simulations between 0 to $\sim35$ (see right y-axis in Fig. \ref{Spectrum_Rcz}), one obtains very similar spectrum for the four simulations. All numerical models show a significant amount of energy for frequencies up to $\omega/\omega_{\rm conv} \sim 5$ and for harmonic degree $\ell$ between 0 and 100.
It is thus interesting to find that a proper rescaling can provide similar power spectra independently of the luminosity enhancement factor.
But  the frequency range is very different for each simulation, with a shift towards higher frequency for larger luminosity enhancement factors.
The normalised frequency value $\omega/\omega_{\rm conv} \sim 5$ corresponds to $\sim \uHz{6}$ for the \textit{ref} simulation,  $\sim \uHz{14}$ for \textit{boost1d1}, $\sim \uHz{30}$ for \textit{boost1d2} and $\sim \uHz{145}$ for \textit{boost1d4}. Thus, the higher the boost, the larger the energy in convective eddies of high frequencies.

\subsection{Radiative zone}
\label{sec:Ek_RZ}

\begin{figure*}
\centering
   \includegraphics[width=0.49\textwidth]{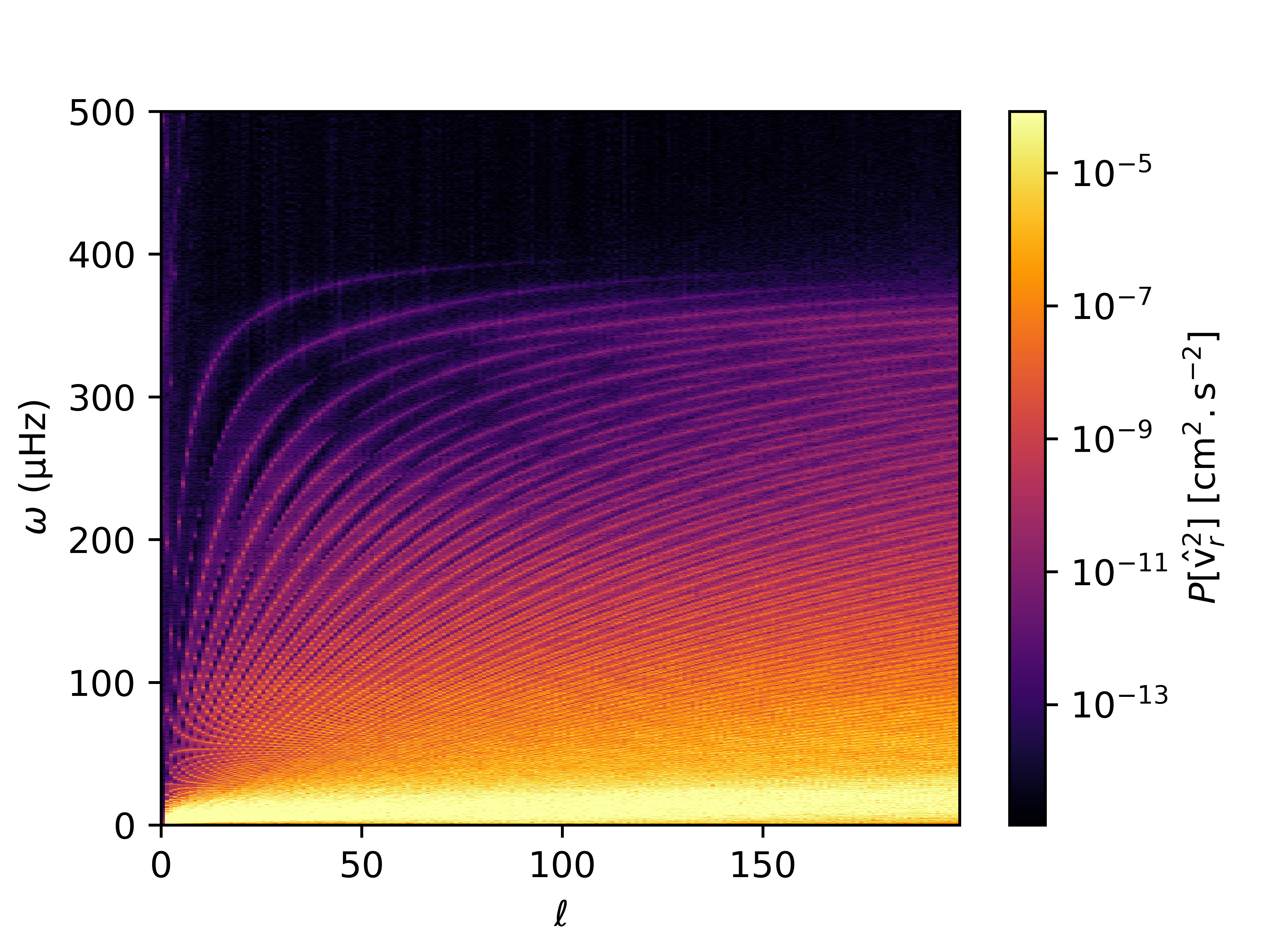}
   \includegraphics[width=0.49\textwidth]{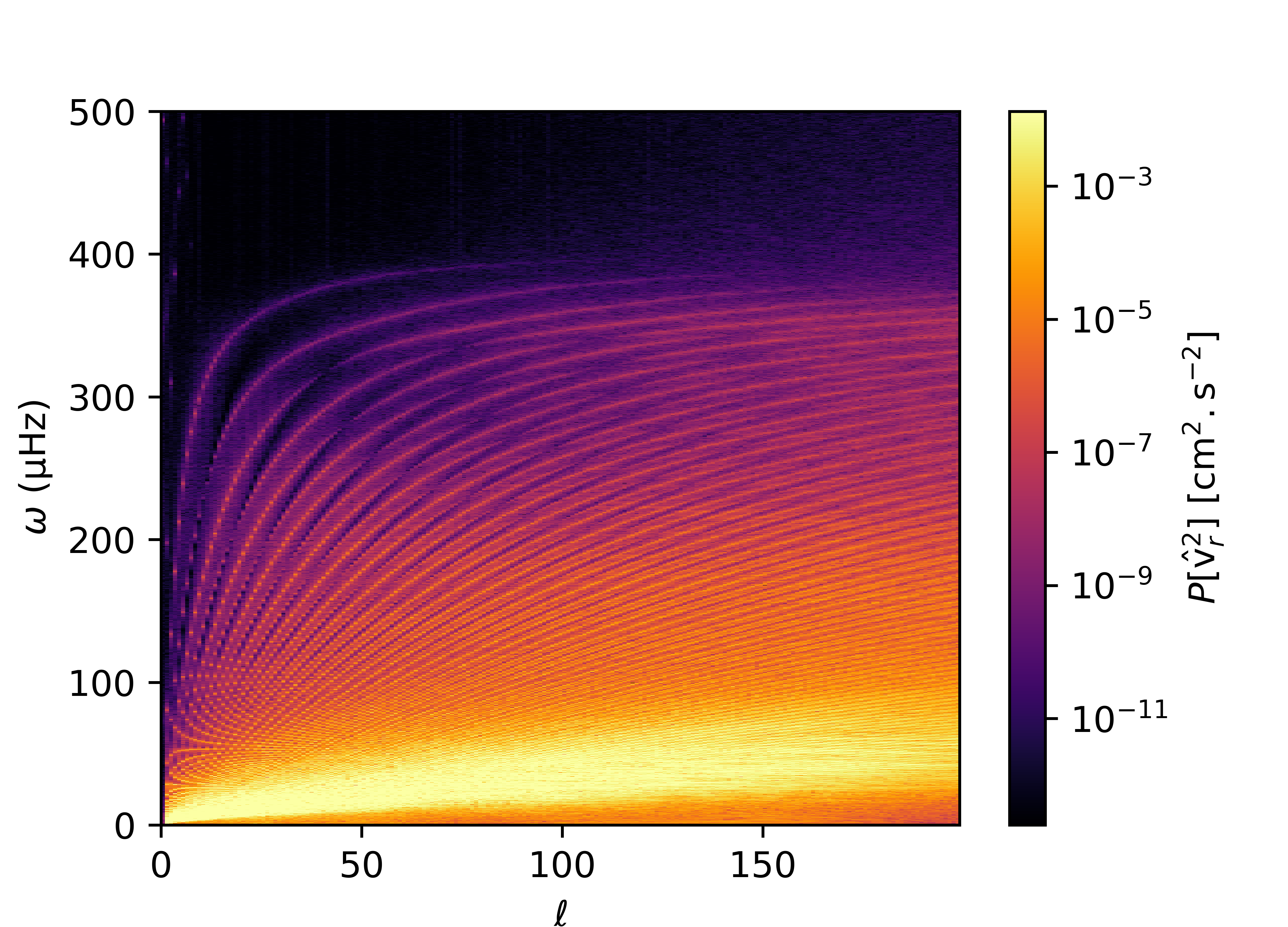}
   \includegraphics[width=0.49\textwidth]{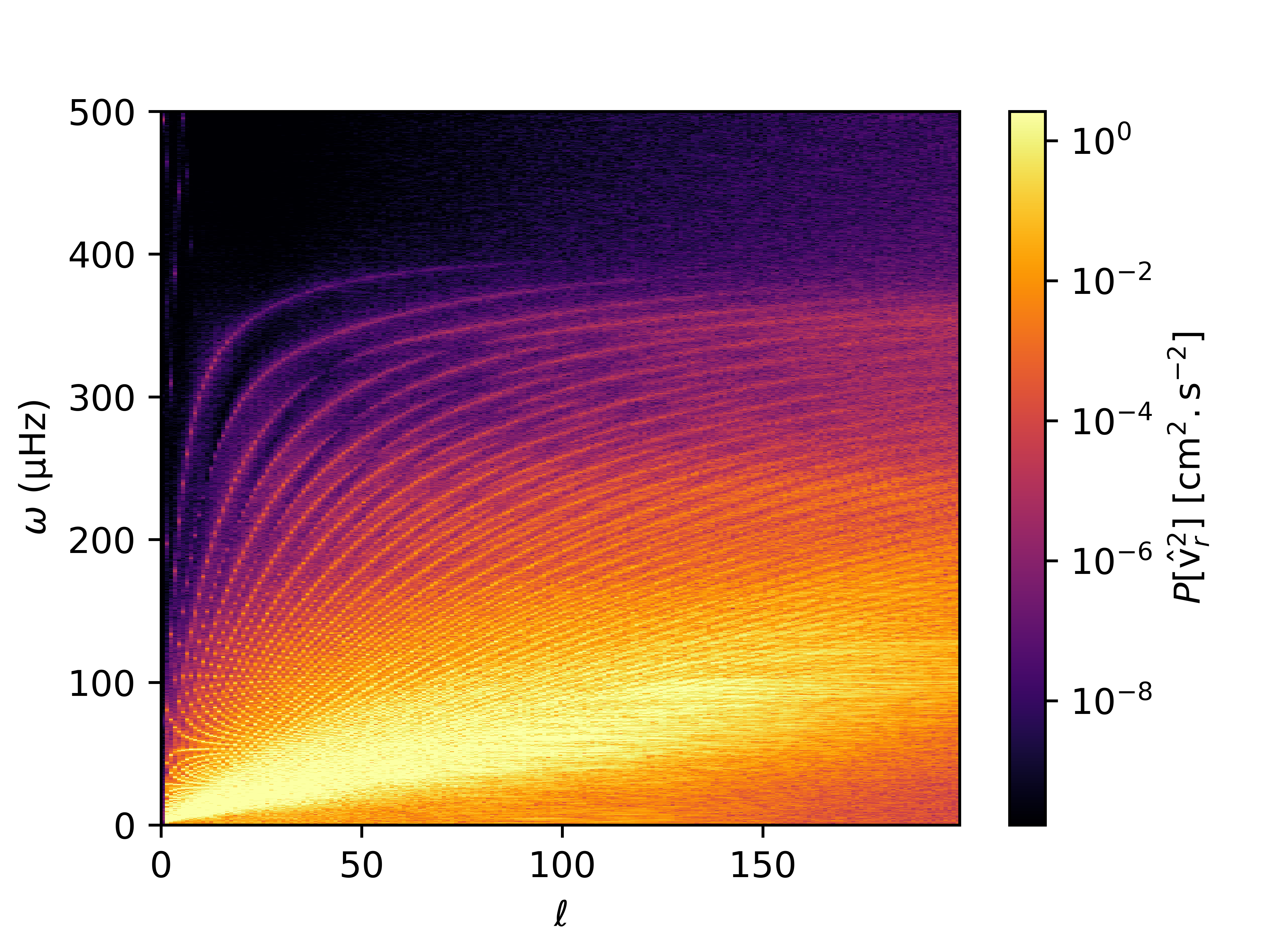}
   \includegraphics[width=0.49\textwidth]{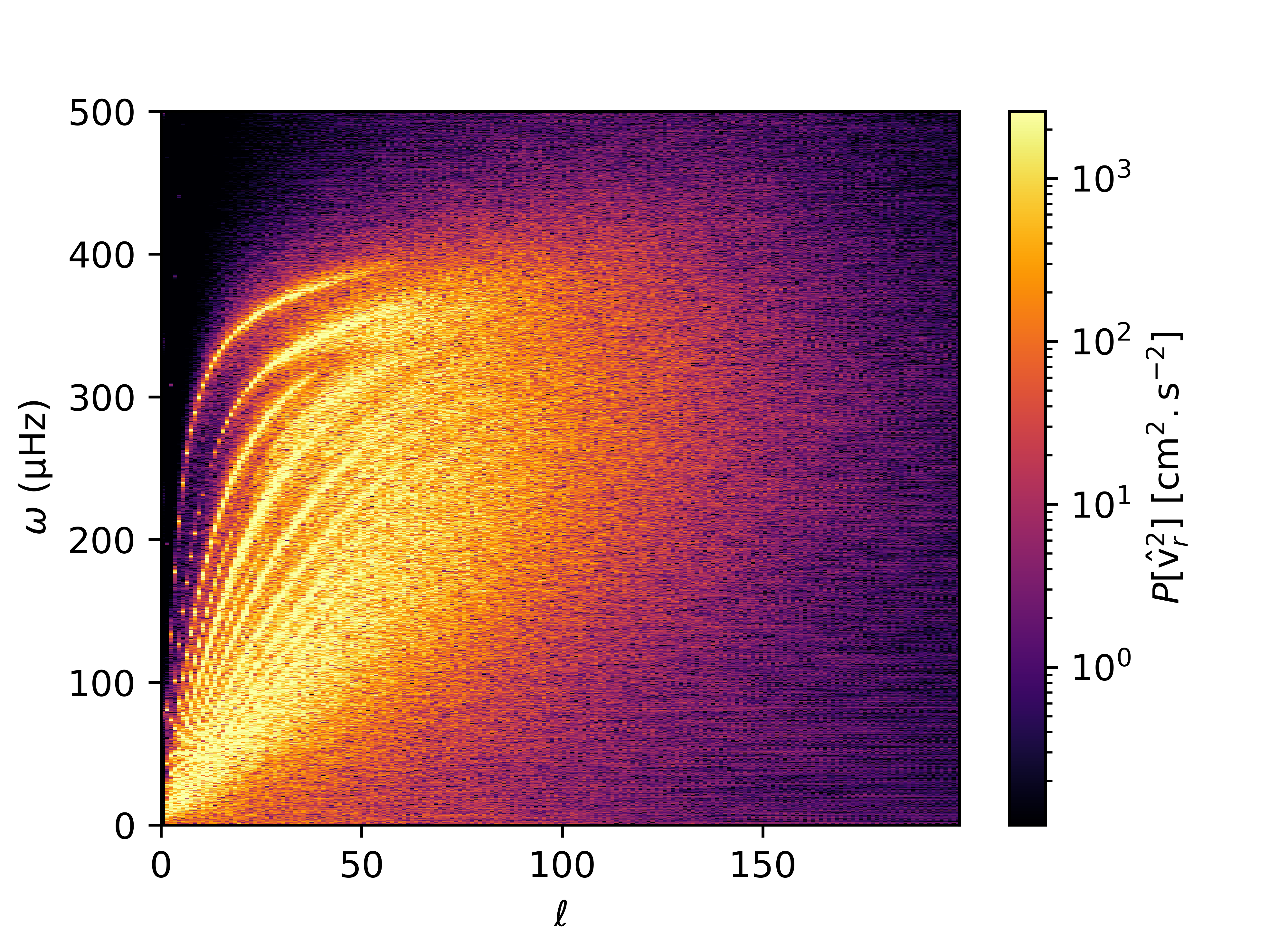}
     \caption{Power spectra of the radial velocity for the four stellar simulations \textit{ref} (top left), \textit{boost1d1} (top right), \textit{boost1d2} (bottom left), and \textit{boost1d4} (bottom right) at depth $r=0.494R_{\rm tot} \simeq r_{\rm conv} - 2H_{p,\rm{conv}}$. They were obtained via mode projection on the spherical harmonics basis and a temporal Fourier transform of the radial velocity. Velocities are not rescaled, and the magnitudes represented by the colour bar are different for each simulation.}
     \label{fig:IGW_spectra_RZ}
\end{figure*}

As in Sect. \ref{conv}, we now analyse the power spectra of the radial velocity in the radiative zone.
Figure \ref{fig:IGW_spectra_RZ} shows the power spectra of the radial velocity at radius $r=0.494 R_{\rm tot}$, which is approximately at two pressure scale heights $H_{p,{\rm conv}}$ from the convective boundary.
At such a depth, located far away from the convective boundary, one can reasonably assume that the waves are the main contributor to the velocity and that the contribution from penetrative plumes is negligible. Velocities are not rescaled and the magnitudes represented by the colour bar are different for each simulation in Fig.~\ref{fig:IGW_spectra_RZ}.

A pattern of bright ridges of high energy is present in the four panels of Fig.~\ref{fig:IGW_spectra_RZ}. This structure is similar to the one obtained by linear theory \citep[see for example][]{Christensen14} and by other numerical simulations of solar-like stars \citep{Alvan2014, Alvan2015} and of more massive stars with a convective core \citep{Horst2020}. These bright ridges present a discrete nature, and the observed bright dots correspond to standing gravity waves, or g modes, that form in the radiative zone (see Sect.  \ref{g-modes}). The g-mode patterns have similar structures for the four numerical models, even if they are only visible at low $\ell$ in \textit{boost1d4}.
As will be shown in Sect.  \ref{g-modes}, the eigenfrequencies of the g modes are not affected by the artificial enhancement of the luminosity. These frequencies depend only on the stratification and the geometry of the resonant cavity limited by the inner boundary, located at $r=0.4R_{\rm tot}$ in present simulations, and by the radiative/convective boundary $r_{\rm conv}$.

The power spectra in Fig. \ref{fig:IGW_spectra_RZ} are displayed for frequencies ranging from 0 to $\uHz{500}$ and for degrees $\ell$ from 0 to 200. In order to reach such high frequencies, the simulation data need to be sampled at a fixed time interval that is short enough ($\le10^3$s) to capture the full spectrum of the waves up to the Brunt-Väisälä frequency. According to Fig. \ref{fig:BV}, the maximal frequency for IGWs propagating at radius $r=0.494R_{\rm tot}$ is $\simeq \uHz{370}$. We thus identify the modes observed at low degrees, $\ell \simeq 1, 2$ or $3$, and at frequencies of $\sim \uHz{400}$ or above as standing acoustic waves, or p-modes. The study of p-modes is beyond the scope of this work.

Figure \ref{fig:IGW_spectra_RZ} clearly shows that the larger the luminosity enhancement factor, the higher the energy of high frequency waves. This trend is consistent with theoretical models.
For waves excited by turbulent Reynolds stress, \citet{Lecoanet13} predict a peak of the IGW flux for waves with frequencies close to the convective turnover frequency, $\omega_{\rm conv}$, which increases with the luminosity enhancement factor.
For waves generated by penetrative plumes, \citet{Pincon2016} suggest that when the frequency associated with the lifetime of the plumes increases, there is a redistribution of the wave energy from low frequencies towards higher frequencies. The lifetime of the plumes can be linked to their velocities in the overshooting layer, which significantly increase with the luminosity enhancement factor (see Sect. \ref{sec:rms_vel} and Paper I).

In terms of length scales, Fig. \ref{fig:IGW_spectra_RZ} shows that most of the energy tends to be concentrated in waves of lower degree $\ell$ for increasing luminosity enhancement factors. The higher the enhancement factor is, the lower the energy in waves of small length scales (high $\ell$) compared to the larger ones (low $\ell$). This can also be expected from theory, eddies with characteristic degree $\ell_{\rm eddy}$ will excite waves of degree $\ell \leq \ell_{\rm eddy}$ \citep{Lecoanet13}. This assumption comes from the statistical properties of stellar convection using Kolmogorov turbulence and is often used to model convection in the context of IGW excitation \citep{Stein1967, Goldreich1977, Goldreich1990, Zahn1997}. It can be expressed as
\begin{equation}
    \ell_{\rm eddy} \sim \Lambda \left(\frac{\omega}{\omega_{\rm conv}} \right)^{3/2},
    \label{eq:ell_max}
\end{equation}
where $\Lambda$ corresponds to the size of the largest convective eddies. Based on a comparison of the flow for all simulations, we assume that $\Lambda$ is the same for the four numerical models; thus, the value of $\ell_{\rm eddy}$ for a given frequency is smaller for a more boosted simulation.

In summary, the waves that bear most of the energy occur at different frequency and spatial ranges for each simulation. In the \textit{ref} simulation most of the energy is below $\sim \uHz{30}$ and spread over all $\ell$ (up to 200) while it is above $\sim \uHz{50}$ and for $0 \leq \ell \leq 100$ for the \textit{boost1d4} simulation. In addition to Sect. \ref{conv}, these results confirm the theoretical expectations that convective eddies with higher frequencies excite IGWs with higher frequencies \citep{Kumar1999, Lecoanet13}. This should be kept in mind when studying IGWs, and particularly energy and angular momentum transport, as they strongly depend on the frequencies and angular degrees of the waves that can be excited \cp[see for example ][]{Zahn1997}. 

\subsection{g modes}
\label{g-modes}
In order to confirm that the bright dots patterns of Fig. \ref{fig:IGW_spectra_RZ} are indeed g modes, we compare their frequencies along $\ell$ slices to the results of a linear stability analysis using the stellar oscillation code GYRE (Version 6.0)\footnote{https://gyre.readthedocs.io/} \citep{Townsend2013, Townsend2018}. This code solves the oscillation equations and provides the eigenfrequencies and eigenfunctions characteristic of a one-dimensional stellar structure model. As an input to GYRE, we used the initial one-dimensional radial profile common to all four simulations, with a domain geometry corresponding to the radially truncated domain of our simulations.

\begin{figure}[!h]
    \centering
    \includegraphics[width=0.5\textwidth]{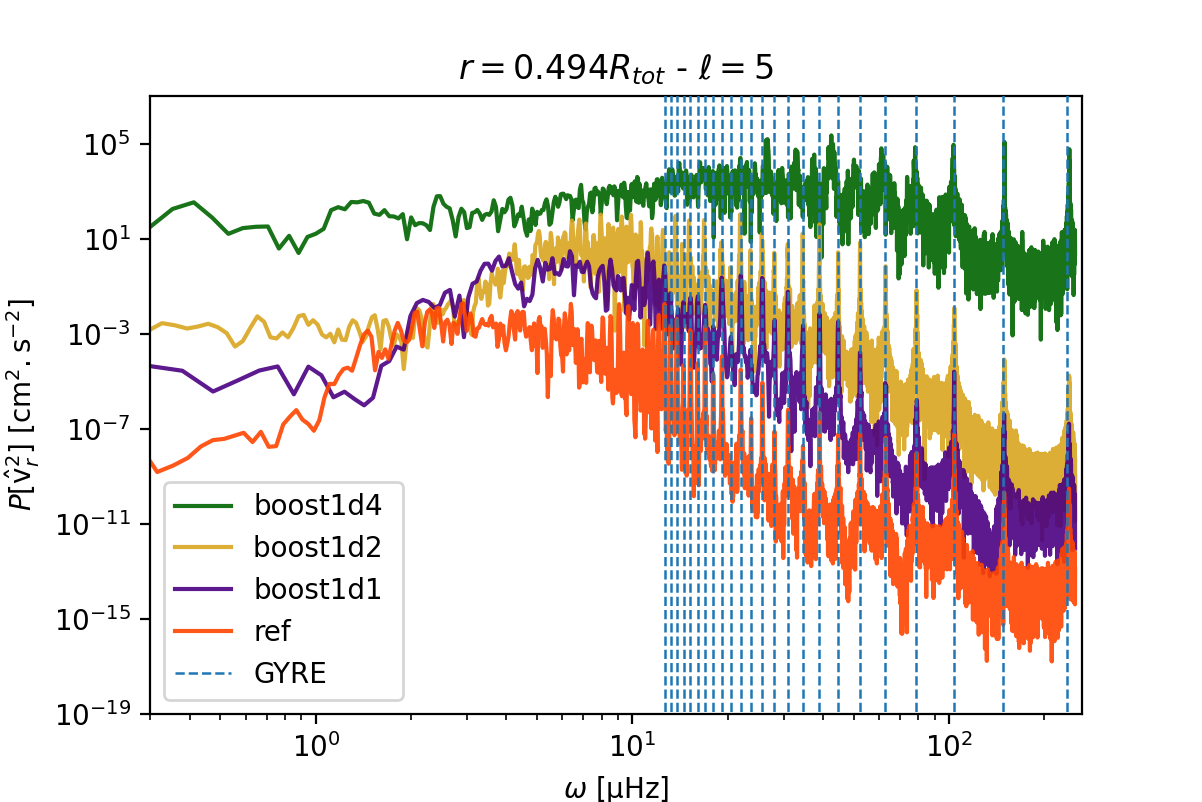}
    \caption{Power spectrum of the radial velocity as a function of frequency at degree $\ell = 5$ for the four simulations \textit{ref}, \textit{boost1d1}, \textit{boost1d2,} and \textit{boost1d4} at depth r = 0.494 $R_{\rm tot} \simeq r_{\rm conv} - 2H_{P,{\rm conv}}$. The vertical dashed blue lines correspond to g-mode frequencies predicted by GYRE. }
    \label{fig:PowervsFreq_GYRE}
\end{figure}

Figure \ref{fig:PowervsFreq_GYRE} shows slices of the spectra of Fig. \ref{fig:IGW_spectra_RZ} for degree $\ell = 5$.
The peaks in the spectra correspond to the bright dots of Fig. \ref{fig:IGW_spectra_RZ}. Those peaks obtained from the four MUSIC simulations closely match the predictions of GYRE (vertical dashed lines) for the g-mode frequencies. For better visibility we do not plot the GYRE predictions for modes with frequency lower than 12$\rm \mu$Hz.
The degree $\ell=5$ is chosen arbitrarily; we also found a good match for the other angular degree we tested, such as $\ell$ = 1, 2, and 10.
For each numerical model, the energy spectrum peaks at a different frequency. For the \textit{ref} simulation, the peak is visually identified at $\simeq \uHz{2}$, for \textit{boost1d1} at $\simeq \uHz{3.5}$, for \textit{boost1d2} at $\simeq \uHz{6}$ and for \textit{boost1d4} at $\simeq \uHz{25}$. These frequencies are rather close to the convective turnover frequency of the corresponding simulation (see Table \ref{tab1}). The location of the maximum power is linked to the excitation of IGWs as analysed in the next section.

%%%%%%%%%%%%%%%%%%%%%
%   SECTION 5
%%%%%%%%%%%%%%%%%%%%%

\section{Excitation and damping of internal gravity waves}
\label{sec:IGW}

\subsection{Radial evolution of power spectra}
\label{sec:radial_spectra}

Next, the aim is to understand the impact of the luminosity enhancement on the excitation of IGW and their propagation in the radiative zone. To do so, we analyse in this section the evolution with depth of the power spectral density (PSD) of the radial velocity summed over the harmonic degree $\ell$, which determines which frequencies contribute the most to the energy.
The relation between the power spectrum, $P[\hat{\vel}_r^2]$, and the PSD, $\mathrm{PSD}[\hat{\vel}_r^2]$, is given in our conventions by
\begin{equation}
    P[\hat{\vel}_r^2] = \frac{\mathrm{PSD}[\hat{\vel}_r^2]}{T_{\rm s}},
    \label{eq:link_PS_PSD}
\end{equation}
with $T_{\rm s}$ the sampling time that corresponds to the total time span used for the computations of the spectra.
In order to compare the amount of power in log-sized, non-uniform frequency bins,
we look at $\mathrm{PSD}_{\ln \omega}$, the PSD in terms of $\ln \omega$, defined by
\begin{equation}
    \mathrm{PSD}_{\ln \omega} \dif \ln \omega =
    \mathrm{PSD} \dif \omega.
\end{equation}

\begin{figure*}[h!]
\centering
   \includegraphics[width=0.49\textwidth]{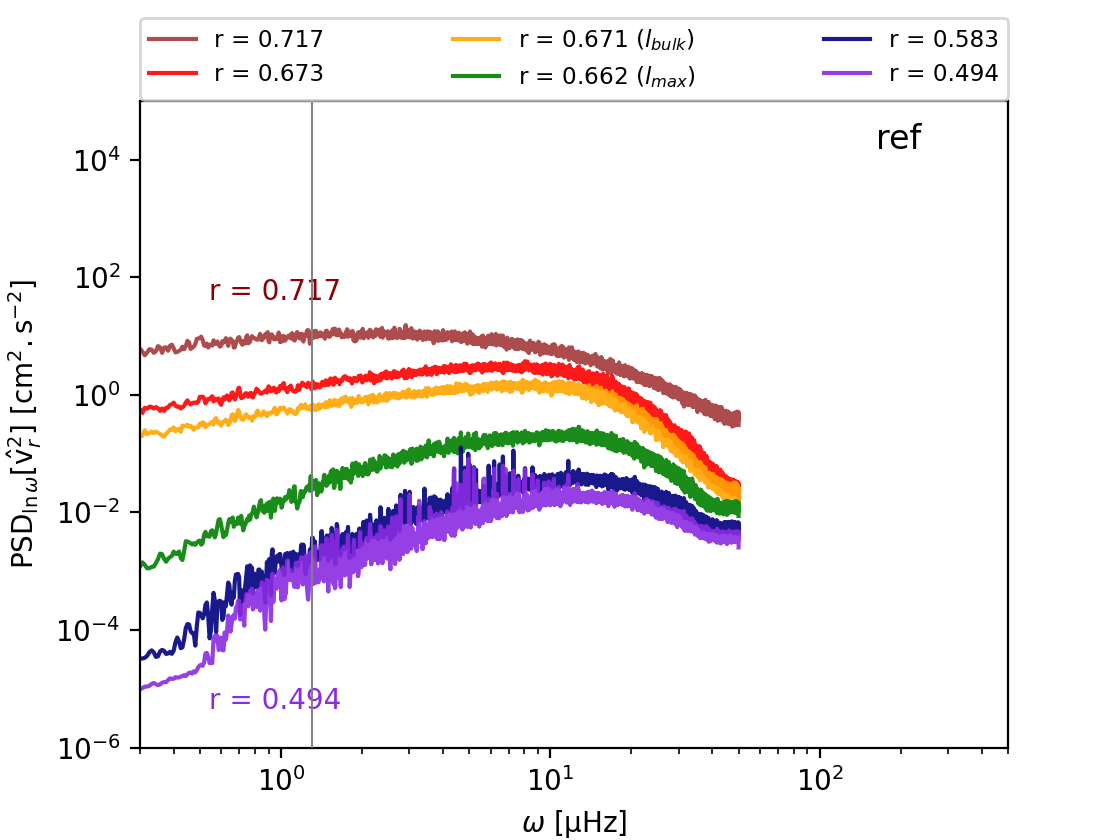}
   \includegraphics[width=0.49\textwidth]{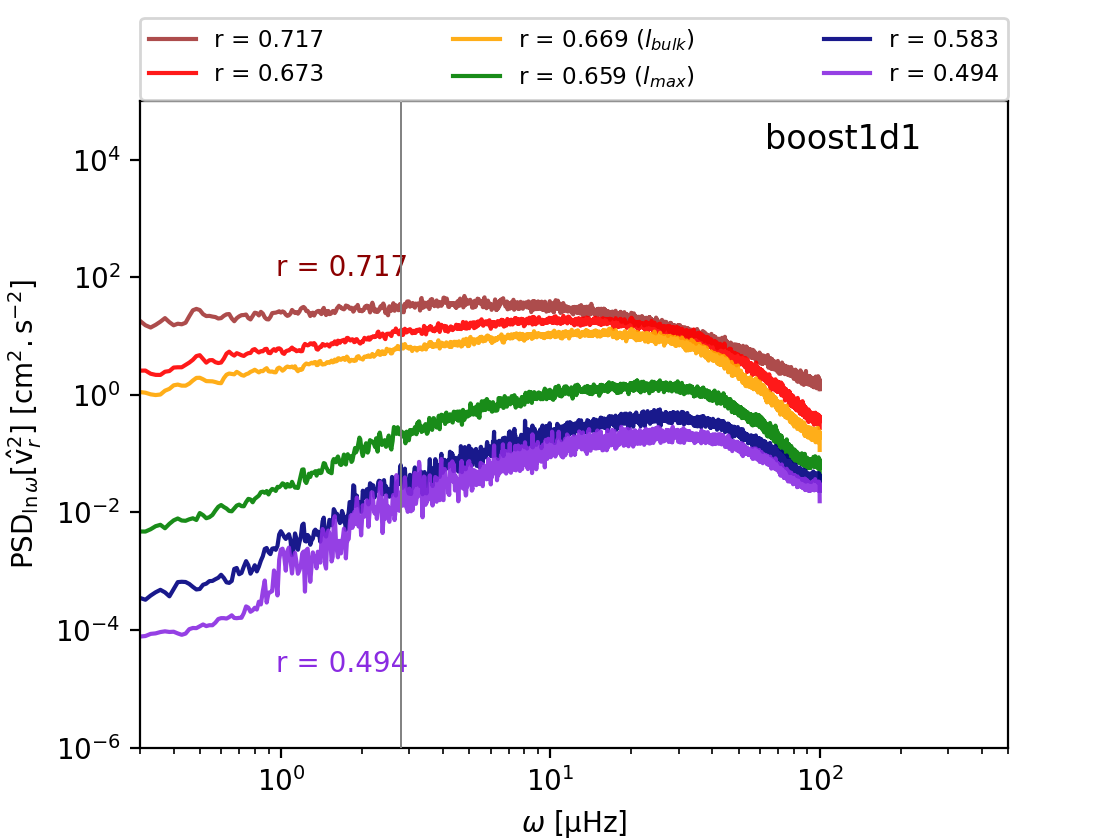}
   \includegraphics[width=0.49\textwidth]{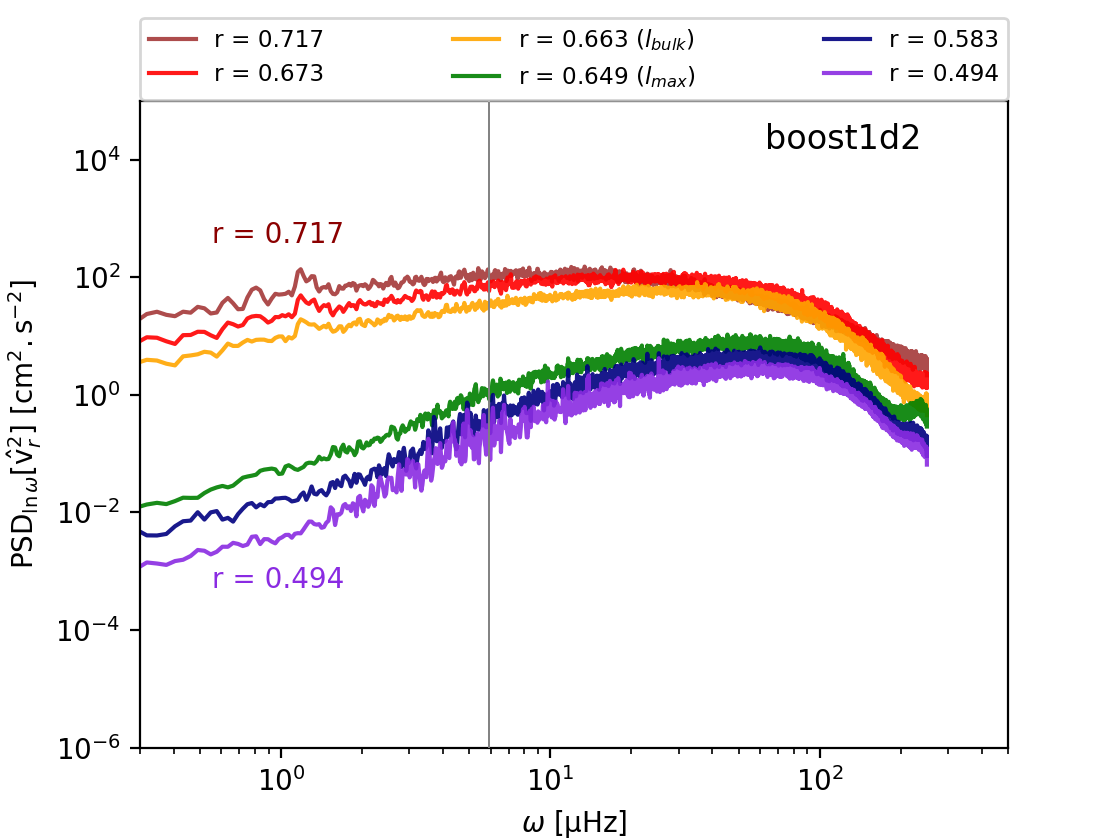}
   \includegraphics[width=0.49\textwidth]{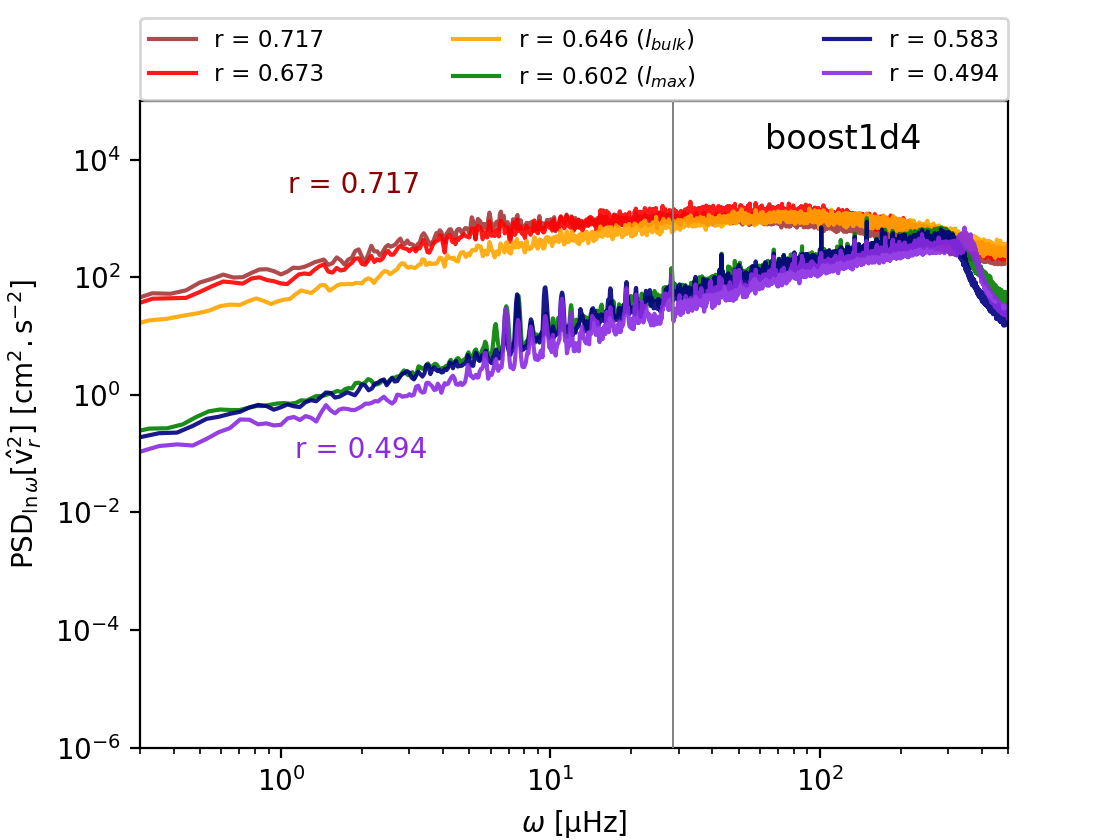}
     \caption{PSD of the radial velocity as a function of frequency for the four simulations \textit{ref} (top left), \textit{boost1d1} (top right), \textit{boost1d2} (bottom left), and \textit{boost1d4} (bottom right) at different depths. The vertical grey line indicates the convective frequency for each simulation. The spectra are obtained via mode projection on the spherical harmonics basis and a temporal Fourier transform of the radial velocity.}
     \label{fig:powerVSfreq_radial_evol}
\end{figure*}

Figure \ref{fig:powerVSfreq_radial_evol} shows
\begin{equation}
    \mathrm{PSD}_{\ln \omega}[\hat{\vel}_r^2](r,\omega) = \sum_{\ell} \omega \,\mathrm{PSD}[\hat{\vel}_r^2](r,\ell,\omega)
    \label{eq:PSD_log_omega}
\end{equation}
at selected radii in the convective and radiative zones for the four simulations.
The spectra are computed over a simulated time of $~200 \times \tau_{\rm conv}$ (see Table~\ref{tab0})
for \textit{ref}, \textit{boost1d1} and \textit{boost1d2} and $~500 \times \tau_{\rm conv}$ for \textit{boost1d4}. This is because $\tau_{\rm conv}$ is short for simulation \textit{boost1d4} compared to other simulations.
For each simulation the frequency range covered by the spectra are different as we focus on the range that bears most of the energy (see Sect. \ref{sec:Ek_RZ}).
Moreover, the flattening of the slope of the spectra close to the maximal frequency for each simulation results from numerical aliasing.
This does not impact our analysis, as we focus on frequencies below the affected range.

We show in Fig. \ref{fig:powerVSfreq_radial_evol} that in the bulk of the convective zone, at $r=0.717 R_{\rm tot} \simeq r_{\rm conv}+H_{p,{\rm conv}}$ (brown curve),
the energy of all four simulations is increasing with frequency up to approximately the convective frequency, $\omega_{\rm conv}$, which is indicated by the vertical grey line. Then, the energy decreases towards higher frequencies, consistent with the results presented in Sect. \ref{sec:CZ}.
Just below the convective boundary, velocities are the result of a mix of convective penetration and waves.
In Paper I, we define a layer of characteristic length $l_{\rm bulk}$ as the distance that convective plumes typically penetrate, and a larger penetration length $l_{\rm max}$ characterised by the most vigorous convective plumes.
The values of $l_{\rm bulk}$ and $l_{\rm max}$ derived in Paper I and used in this work are given in Table \ref{tab:lmax}.
Figure \ref{fig:powerVSfreq_radial_evol} shows spectra in the radiative - convective transition region, at $r=0.673 R_{\rm tot}$ (red curve), and at $r=r_{\rm conv}-l_{\rm bulk}$ (yellow curve) for all simulations.

We first discuss simulations $ref$, $boost1d1,$ and $boost1d2$. Compared to spectra above the convective boundary, the energy increases up to some characteristic frequency and sharply drops beyond.
As the luminosity enhancement factor increases, the knee of the spectrum shifts to higher frequencies;
in addition, the spectra above and just below the convective boundary (brown, red, yellow curves) become closer to each other,
indicating that vigorous convection increasingly dominates at the top of the penetration region.

Finally, at even deeper layers inside the radiative zone, the PSDs are dominated by waves,
excited from above and propagating towards the centre of the star
while also undergoing damping \citep{Press1981}, which we analyse in Sect. \ref{sec:spatial_damping}.
The deepest spectra displayed in Fig.~\ref{fig:powerVSfreq_radial_evol} correspond to the depths
$r=0.583 R_{\rm tot} \simeq r_{\rm conv}-H_{p,{\rm conv}}$ (blue) and $r=0.494 R_{\rm tot} \simeq r_{\rm conv}-2H_{P,{\rm conv}}$ (purple curve).
In this region, the spectra present high-amplitude g modes.
They are also decreasing towards low frequencies, reaching a peak for some $\omega$,  which corresponds approximately to the frequency at which the spectra reach their maxima just below the convective boundary (red and yellow curves). However, at low frequency we can observe a modification of the aspect of the spectra.
In this range there are no more g modes, due to the stronger damping of low-$\omega$ waves (see Sect. \ref{sec:spatial_damping}).

The PSD computed at a distance $l_{\rm max}$ from the convective boundary (green curve) is a mix of the spectra just below the convective boundary (red and yellow) and the ones deeper in the radiative zone (blue and purple). Indeed, very little of the convective motions penetrate that deep in the radiative zone, and below a depth of $l_{\rm max}$, the wave signal dominates the spectra. \\

In Fig.~\ref{fig:powerVSfreq_radial_evol}, the most boosted numerical model \textit{boost1d4} is qualitatively different and deserves special discussion.
Firstly, its spectra just below the convective boundary (red, yellow) are very similar to the spectrum in the convection zone (brown).
This could result from the layer becoming very close to convective below the boundary:
as noted in Paper I, the thermal background in the overshooting layer of this model is significantly modified during the course of the simulation,
with the temperature profile getting steeper and closer to the adiabatic gradient.
Secondly, in the region $r \leq r_{\rm conv} - l_{\rm max}$ of the radiative zone where only waves remain,
the spectra are monotonically increasing, compared to their less boosted counterparts, up until the Brunt-Väisälä frequency ($N \simeq \uHz{370}$ at $r = 0.494 R_{\rm tot}$),
after which the energy drops suddenly, since no IGWs can propagate above this frequency.
The energy spectra are therefore `clipped' at high frequencies $\omega \geq N$, resulting in a significant redistribution of wave energies.

\begin{table}[t]
   \caption{Characteristics lengths $l_{\rm bulk}$  and $l_{\rm max}$ derived in Paper I.}
   \label{tab:lmax}
   \centering
   \begin{tabular}{l c c c c}
     \hline \hline
     Simulation &  $l_{\rm bulk}/R_{\rm tot}$ &  $l_{\rm bulk}/H_{p,{\rm conv}}$ &  $l_{\rm max}/R_{\rm tot}$ &  $l_{\rm max}/H_{p,{\rm conv}}$  \\
      \hline
      ref &  $2.35 \times 10^{-3}$ &  $2.62 \times 10^{-2}$ & $1.12 \times 10^{-2}$ & 0.124 \\
       boost1d1 &  $4.72 \times 10^{-3}$ &  $5.24 \times 10^{-2}$ & $1.45 \times 10^{-2}$ & $0.161$ \\
        boost1d2 &  $1.04 \times 10^{-2}$ & $0.115$ & $2.47 \times 10^{-2}$ & $0.275$\\
        boost1d4 &  $2.82 \times 10^{-2}$ &  $0.313$ & $7.1 \times 10^{-2}$ & $0.787$\\
      \hline
   \end{tabular}
   \tablefoot{Values of $l_{\rm bulk}$  and $l_{\rm max}$ based on the vertical heat flux.}
\end{table}

\begin{figure*}
\centering
   \includegraphics[width=0.49\textwidth]{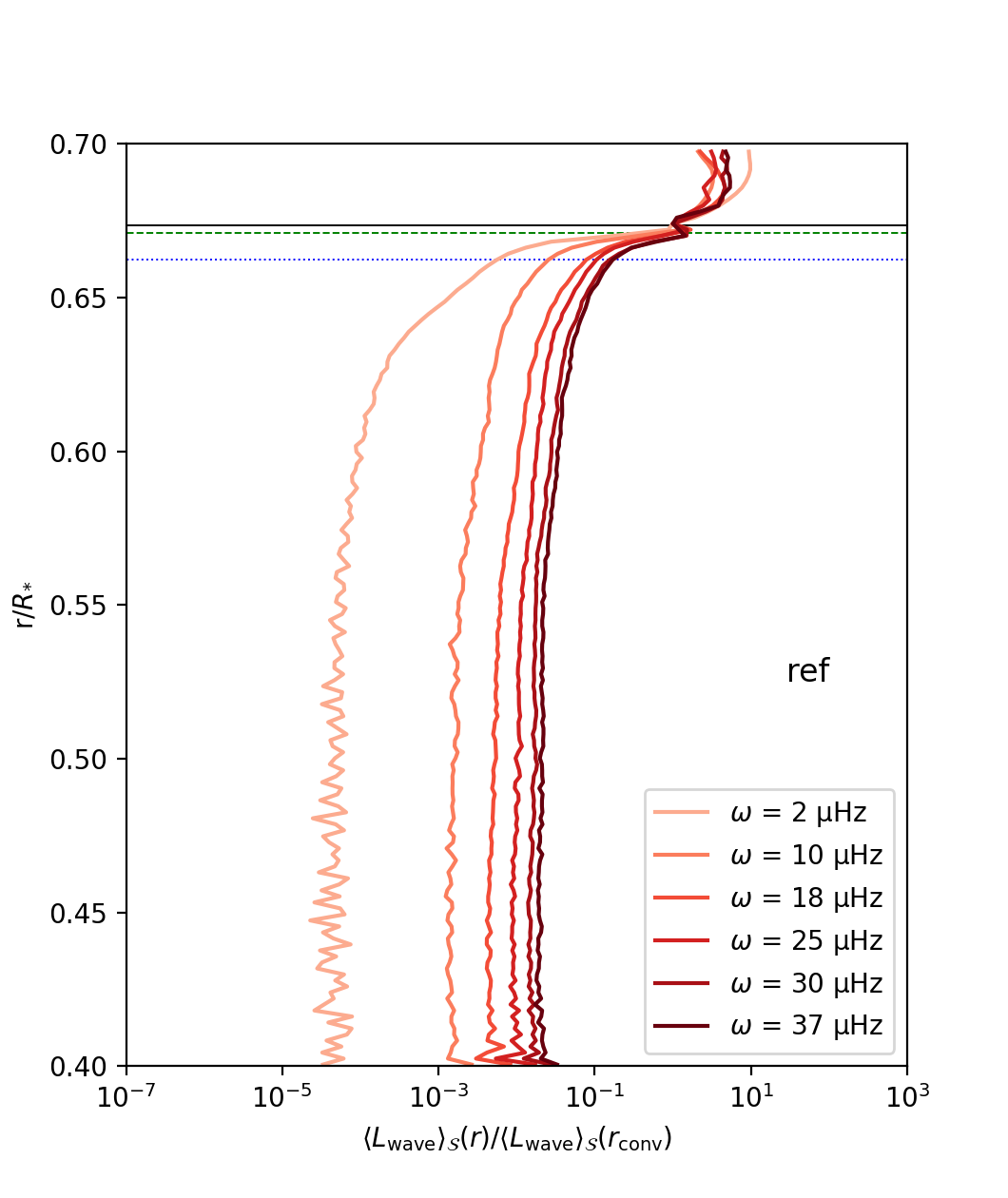}
   \includegraphics[width=0.49\textwidth]{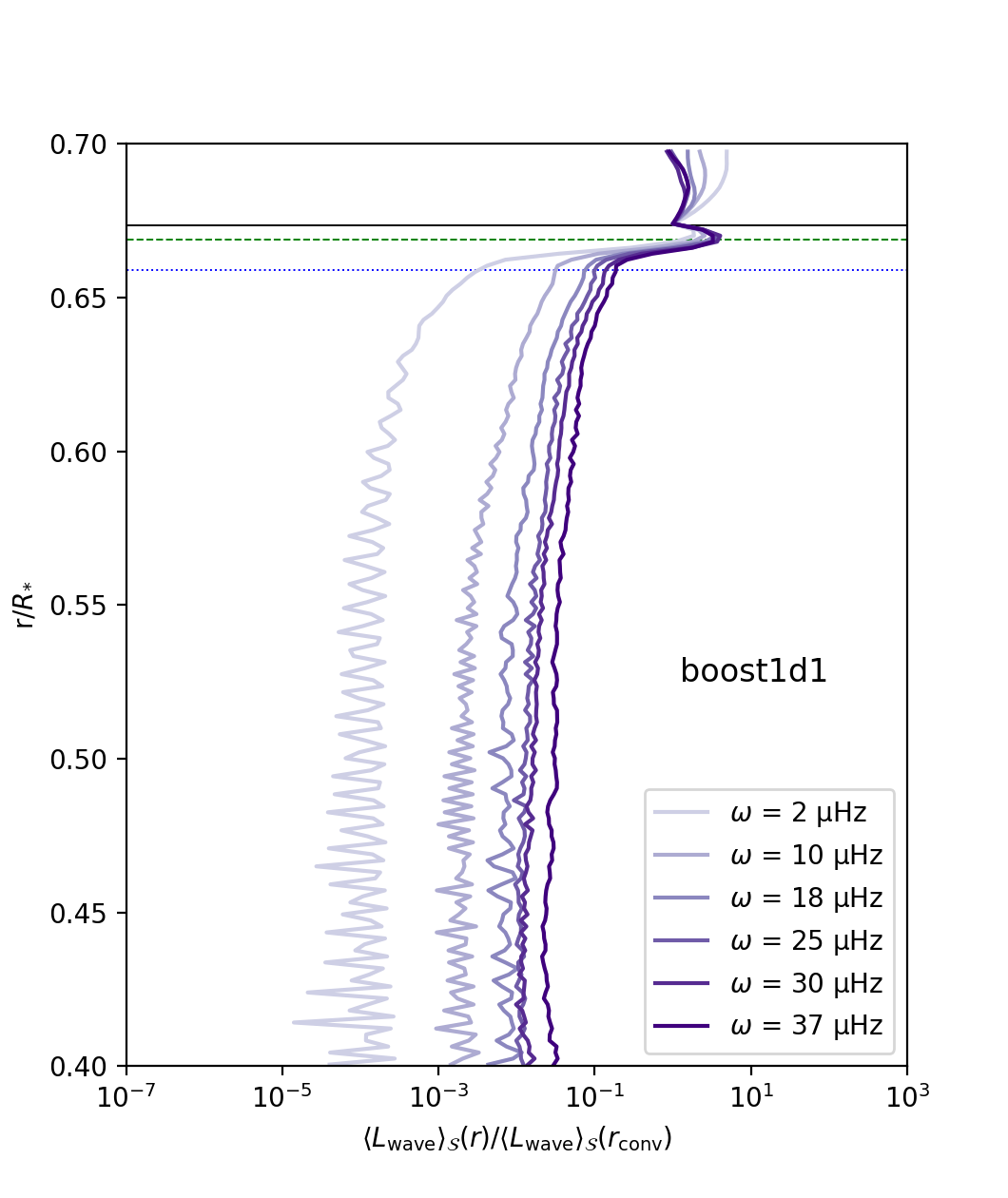}
   \includegraphics[width=0.49\textwidth]{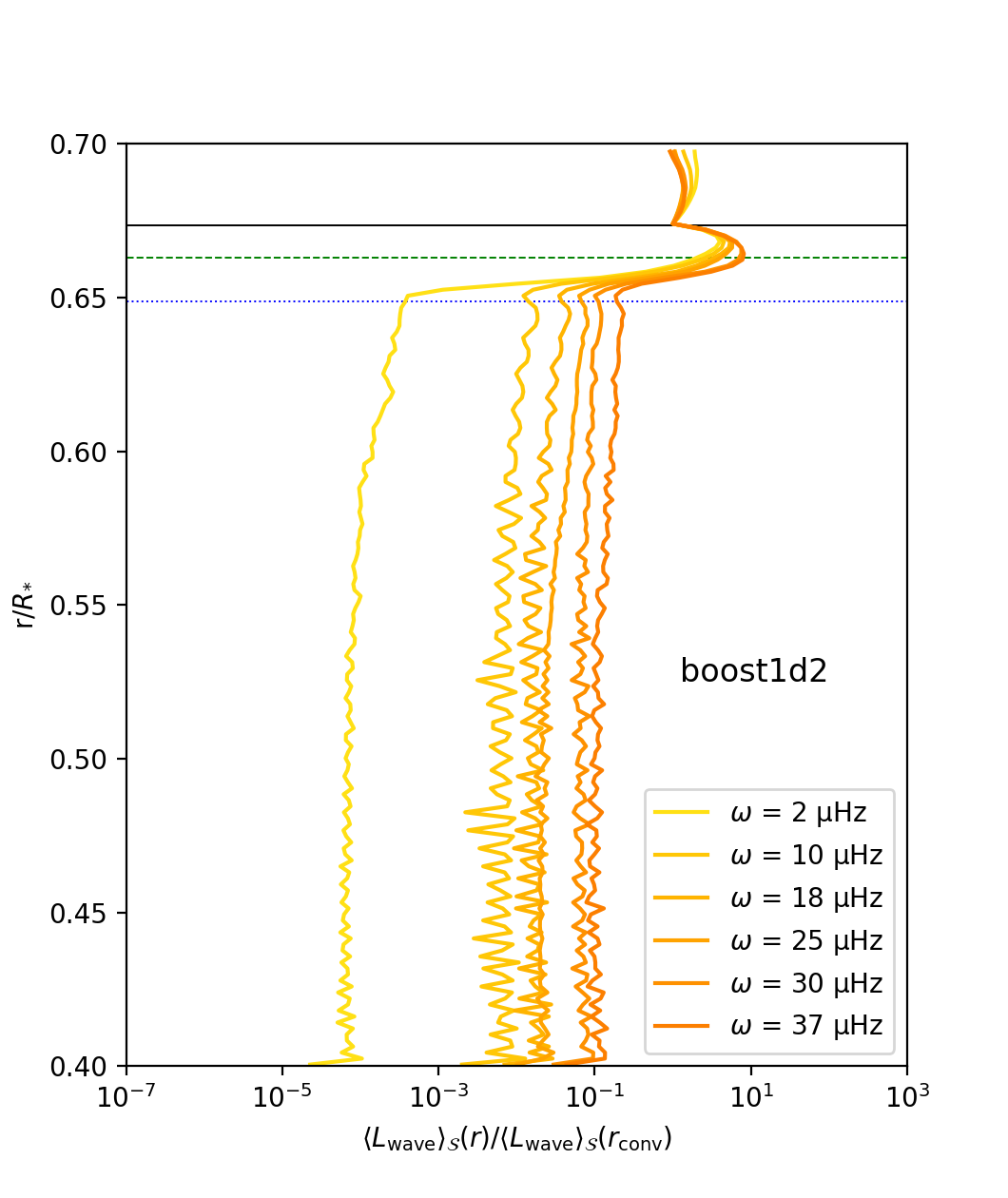}
   \includegraphics[width=0.49\textwidth]{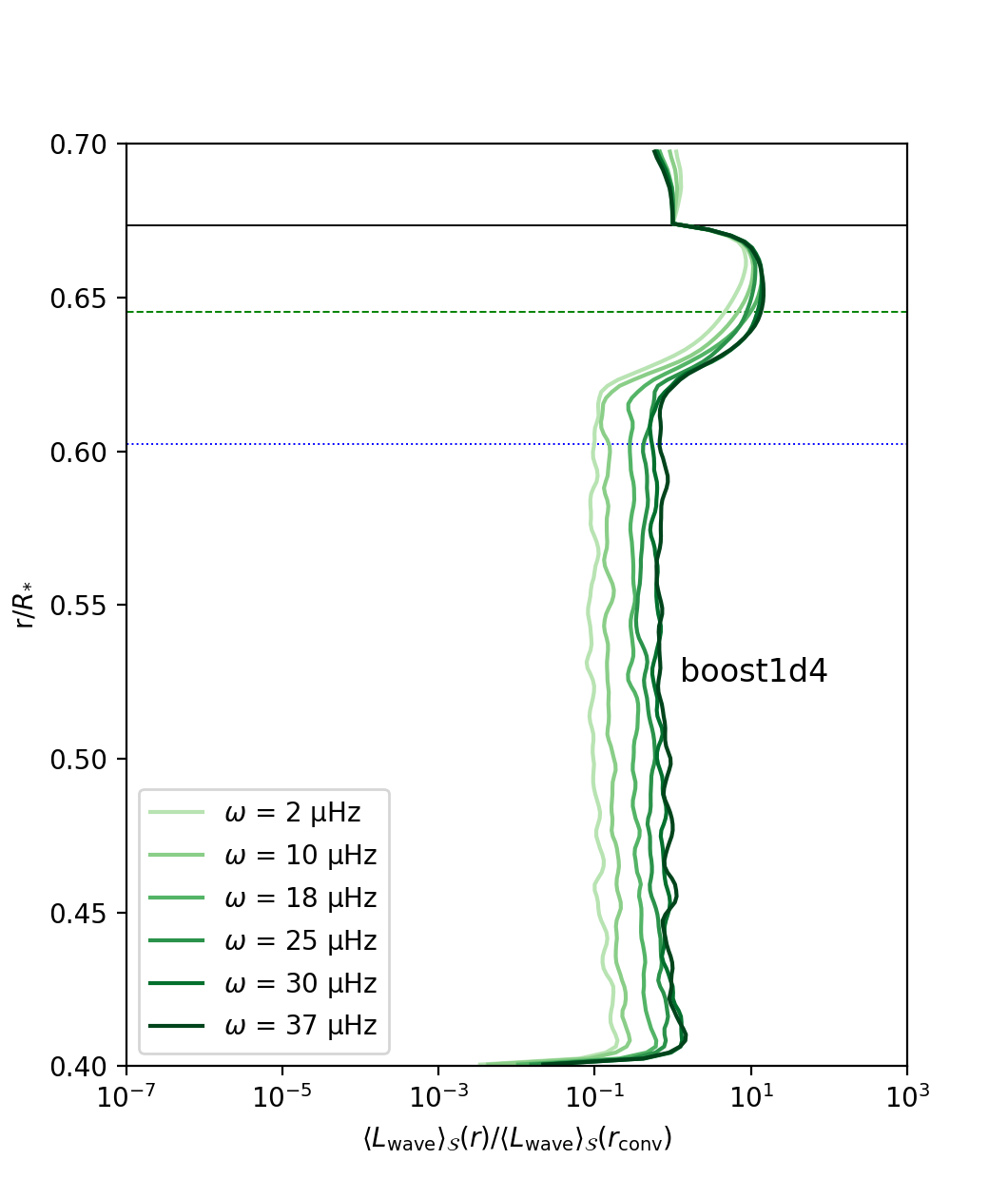}
     \caption{Radial profile of the horizontally averaged luminosity associated with the waves for the four simulations \textit{ref} (top left), \textit{boost1d1} (top right), \textit{boost1d2} (bottom left), and \textit{boost1d4} (bottom right) at six different frequencies. The power spectra are normalised by their value at the convective boundary, $L_{\rm wave}(r_{\rm conv})$. The horizontal solid black line indicates the radiative--convective boundary as defined by the Schwarzschild criterion for the initial model. The horizontal dashed green and dotted blue lines indicate the radii $r=r_{\rm conv}-l_{\rm bulk}$  and $r=r_{\rm conv}-l_{\rm max}$, respectively. The horizontal average, $\left<.\right>_{\mathcal{S}}$, is defined in Eq. \eqref{eq:angular_av}.}
     \label{fig:damping_normCB}
\end{figure*}

\subsection{Amplitude of oscillatory motions across the convective boundary.}

In this subsection, we analyse the decrease in the radial velocity amplitude between the convective zone and the radiative zone.
To this end, we study the IGW wave luminosity as a function of depth.
The luminosity of a single IGW mode is defined as
\begin{equation}
     L_{\rm wave}(r,\ell,\omega) = 4\pi r^2  \Fsingle (r,\ell,\omega),
    \label{eq:lwave}
\end{equation}
with $\Fsingle$ the wave energy flux for the mode. In order to calculate wave fluxes from our numerical simulations and compare them to theoretical predictions,
we estimate the wave energy flux for an individual IGW mode ($\omega$, $\ell$) as \citep[see][]{Press1981, Lecoanet13}
\begin{equation}
\Fsingle  \sim \rho |\hat{\vel}_{\rm h}|^2 u_{\mathrm{g},r},
\end{equation}
where we have assumed that $\vel_{\rm h} \gg \vel_r$ in the radiative zone. We recall that in our two-dimensional case, we identify $\vel_{\rm h} = \vel_{\theta}$. We have introduced $u_{g,r}$, the radial component of the group velocity of the waves, which can be expressed as \citep{Unno1989}
\begin{equation}
    u_{\mathrm{g},r} \simeq \frac{\omega^2}{Nk_{\rm h}}.
\end{equation}
Moreover from \citet{Press1981}, we have
\begin{equation}
    \hat{\vel}_{\rm h} \simeq \frac{N}{\omega}\hat{\vel}_r.
\end{equation}
The  wave flux is thus
\begin{equation}
 \Fsingle \sim \rho \frac{N}{k_{\rm h}}|\hat{\vel}_r|^2.
    \label{eq:flux_real_space}
\end{equation}
Following our definition Eq. \eqref{eq:def_power_spectrum} of the power spectrum
\begin{equation}
    |\hat{\vel}_r|^2 \sim \frac{P[\hat{\vel}_r^2]}{2},
\end{equation}
our final expression for the flux for an individual IGW mode ($\omega$, $\ell$) at radius $r$ is
\begin{equation}
\Fsingle(r,\ell,\omega) \sim \frac{1}{2}\rho \frac{N}{k_{\rm h}} P[\hat{\vel}_r^2](r,\ell,\omega).
    \label{eq:wave_flux}
\end{equation}
The horizontally averaged luminosity of a superposition of modes, Eq. \eqref{eq:lwave}, is therefore given, through Eq. \eqref{eq:wave_flux}, by the total power summed across all $\ell$.

Figure \ref{fig:damping_normCB} shows the radial evolution of the horizontally averaged total IGW luminosity at six different frequencies for the four simulations.
The luminosities are normalised by their value at the convective boundary (horizontal solid black line),
allowing us to compare the depth dependence across the different frequencies.
In absence of damping effects, the wave luminosity of propagating IGWs is conserved,
making it a useful quantity for studying the decay of the wave amplitudes
as they travel inwards, away from the convective boundary.

All four numerical models show a peak of the wave luminosity at $r \sim r_{\rm conv}-l_{\rm bulk}$ (horizontal green dashed line), which is located deeper towards the centre as the luminosity of the simulation increases. We suggest that the peaks indicate the region where the excitation of the waves is maximum.
Below $r \sim r_{\rm conv}-l_{\rm bulk}$ there is a strong decrease in energy towards the centre of the star.
This drop is the result of the transition from convective motions to waves,
in which the majority of the convective kinetic energy is not transferred to IGWs but rather to horizontal flows and to local heating.

After reaching approximately $r\sim r_{\rm conv}-l_{\rm max}$ (horizontal dotted blue line), the energy is still decreasing but at slower rate over some distance for \textit{ref} and \textit{boost1d1}. Below this radius, the wave luminosity is approximately constant. In the case of \textit{boost1d2} and \textit{boost1d4} the wave luminosity is constant from $r\sim r_{\rm conv}-l_{\rm max}$.
%%
%% --------------------------------------------------------------------

The behaviour of the wave luminosity just below the convective boundary shows that the larger the model luminosity enhancement factor, the more energy is transmitted to the waves.
The two expected wave excitation mechanisms, namely turbulent Reynolds stress at the convective boundary and penetrating plumes below the boundary,
directly depend on the model luminosity enhancement factor because of the increase in the convective and penetrating plume velocities (see Sect.  \ref{sec:rms_vel}  and Paper I).
Trying to disentangle the impact of one mechanism from the other or to determine if one is dominating over the other is a difficult task, since both are enhanced with the luminosity of the simulation.
But the changes of behaviour of the spectra (see Fig. \ref{fig:powerVSfreq_radial_evol}) and of the rate of energy decrease (see Fig. \ref{fig:damping_normCB}) linked to the positions of $l_{\rm bulk}$,
where the bulk of the plumes penetrate, and of $l_{\rm max}$ reached by the most vigorous plumes,
suggest that convective penetration plays a non negligible role on the energy spectra of IGWs in the radiative zone.

We also note that the efficiency of the decay of the wave luminosity just below $r \sim r_{\rm conv}-l_{\rm bulk}$ depends on the frequency of the waves.
For a given simulation, the decrease in amplitude of lower frequency waves is larger compared to higher frequency waves.
Compared to their initial amount of energy at the convective boundary, low frequency motions thus have less energy as they propagate towards the centre of the star.
However, the wave luminosity at a distance larger than $l_{\rm max}$ from the convective boundary  is approximately constant and thus is not dependent on the frequency. We suggest that the horizontal averaging masks the effect of radiative damping. From theory it is expected that the amplitude of the waves, represented by $P[\hat{\vel}_r^2]$, decreases as they propagate towards the centre of the star. This expected decay of the velocity amplitude is mostly due to the stratification and the geometry of the numerical models \citep{Press1981, Rogers2017, Ratnasingam2019} rather than to radiative effects.
Damping due to radiative effects will be analysed in more detail in Sect. \ref{sec:spatial_damping}.

The situation for \textit{boost1d4} is less clear.
Because of the very vigorous convection in this model,
waves in the radiative zone reach very large amplitudes shifted towards larger $\omega$ (see Fig.~\ref{fig:IGW_spectra_RZ}),
resulting in significant reflections at the bottom boundary with almost no visible damping;
the problem is further aggravated by the larger overshooting depth of this model.
Waves can also reach strongly non-linear regimes where linear theory breaks down.
The comparison to the other models highlights the limitations of strongly boosted simulations for the analysis of IGW properties.

%%%%%%%%%%%%%%%%%%%%%
%   SECTION 6
%%%%%%%%%%%%%%%%%%%%%

\section{Comparison with theory}
\label{sec:theory}

\subsection{IGW radial energy flux}
\label{radial_flux}
The wave energy flux is central to determining the efficiency of angular momentum transport in stellar interiors, and can be used to predict the detectability of IGWs in stars.
Based on the pioneering work of \citet{Stein1967}, several models have been developed for the calculation of the wave energy flux,
assuming that the excitation mechanism is due to Reynolds stresses but these models result in differences in the predicted flux \citep[e.g.][]{Goldreich1990, Goldreich1994, Kumar1997, Zahn1997, Kumar1999}.
In a recent work,  \citet{Lecoanet13} extend the work of \ct{Goldreich1990} and provide a comparison with the results of previous models.  They analyse the impact of the transition between the convective and radiative regions on the wave energy flux. They characterise the transition by the width of the transition region $d$ and the vertical profile of the buoyancy frequency $N$. If $d$ is small, the waves see a discontinuous transition. In this case,  \citet{Lecoanet13} predict a scaling for the flux (see their Eq.~(40)):
\begin{equation}
\frac{\dif F^{\rm D}}{\dif \ln \omega \dif \ln k_{\rm h}} \propto  k_h^4 \omega^{-13/2}.
\label{eq:fluxD_L13}
\end{equation}
This spectrum was recovered numerically in two-dimensional \citep{Lecoanet2021} and three-dimensional \citep{Couston2018} Cartesian simulations.

\begin{figure*}
\centering
   \includegraphics[width=0.49\textwidth]{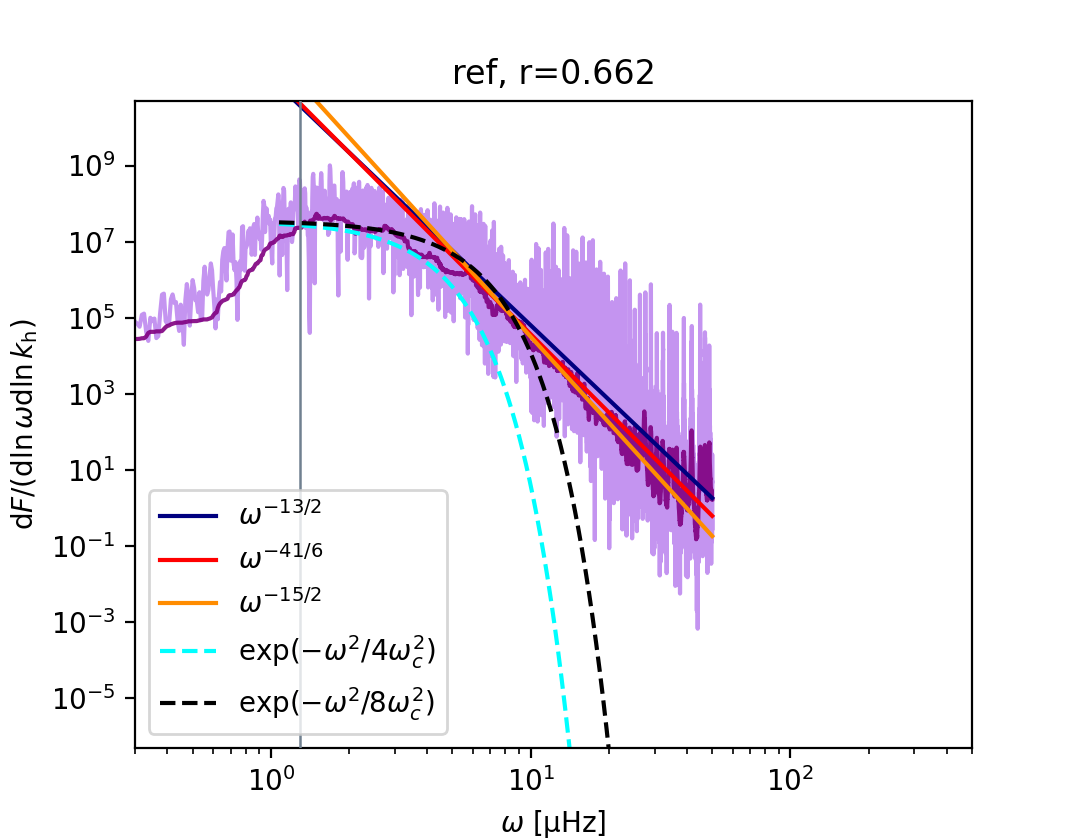}
   \includegraphics[width=0.49\textwidth]{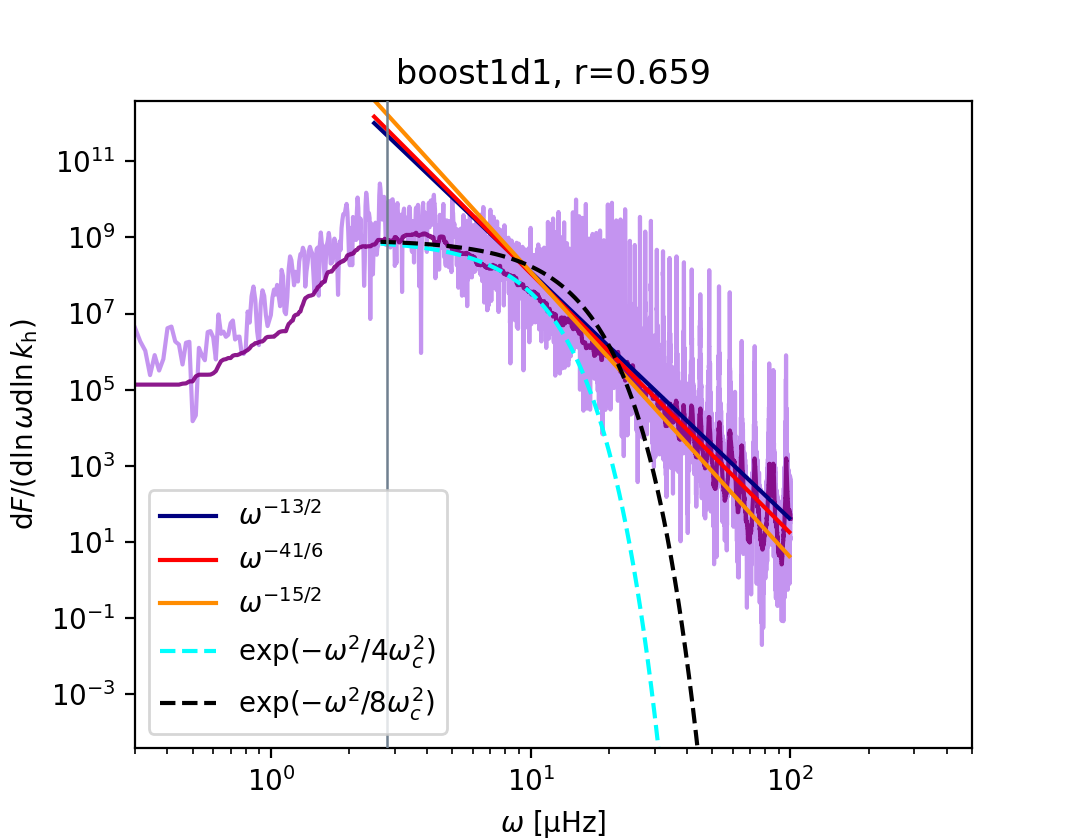}
   \includegraphics[width=0.49\textwidth]{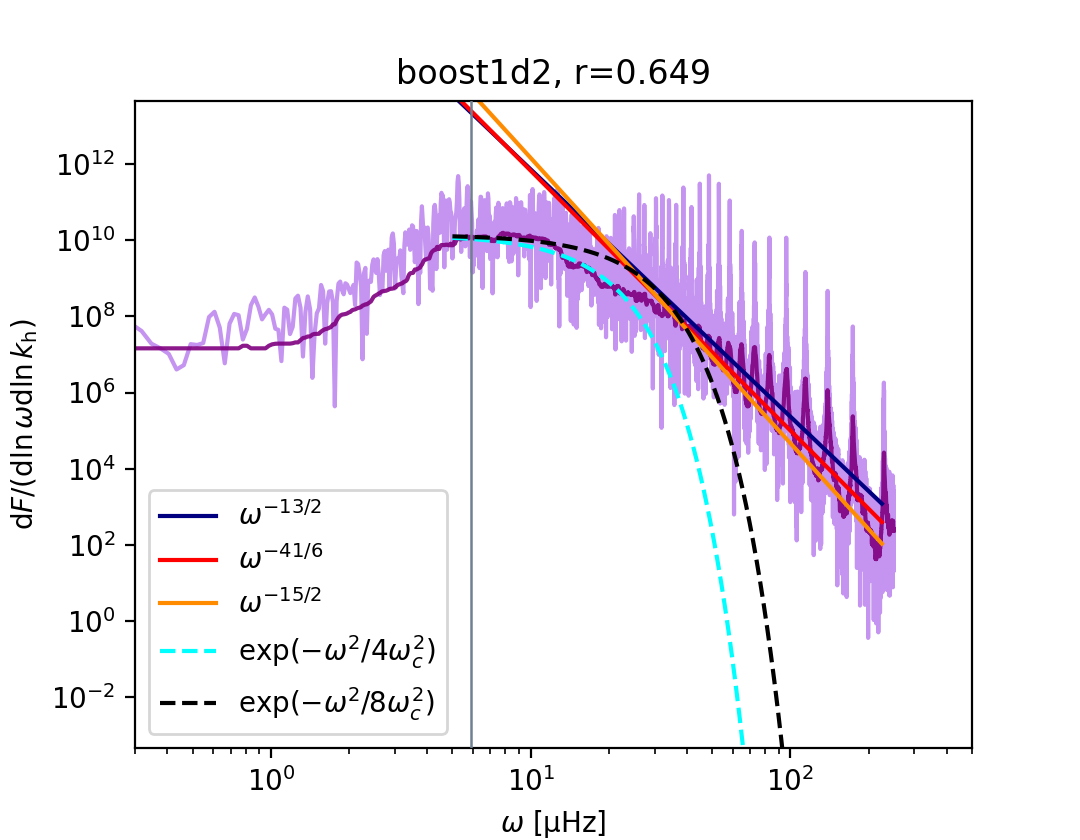}
   \includegraphics[width=0.49\textwidth]{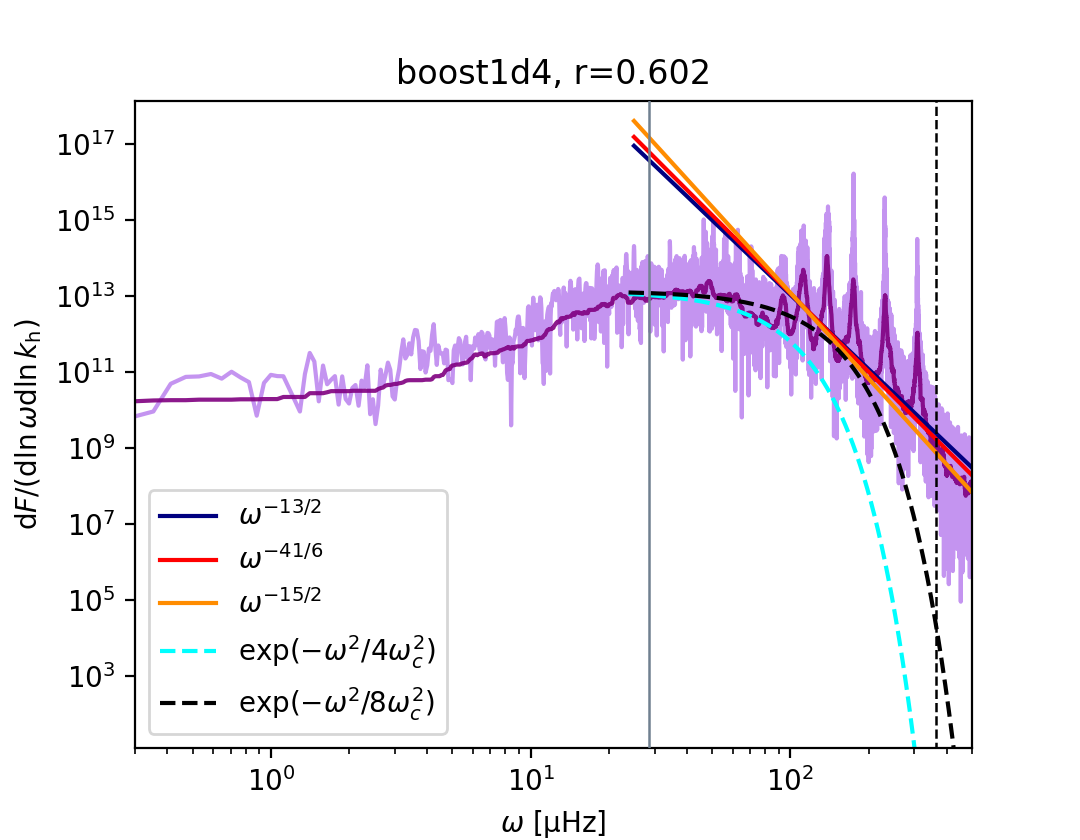}
     \caption{
       Wave energy flux (light purple line) for the four simulations \textit{ref} (top left), \textit{boost1d1} (top right), \textit{boost1d2} (bottom left), and \textit{boost1d4} (bottom right).
       The spectra are computed at $r = r_{\rm conv} - l_{\rm max}$ (i.e. at a distance $l_{max}$ from the convective boundary) and for angular degree $\ell=10$.
       The grey vertical lines indicate the convective turnover frequency, $\omega_{\rm conv}$, and the blue, red, and orange lines correspond to functions with slope  $\omega^{-13/2}$,  $\omega^{-41/6}$, and $\omega^{-15/2}$, respectively.
       The dashed curves represents two Gaussian spectra with characteristic frequencies $\omega_{\rm conv}$ (cyan) and $\sqrt{2}\omega_{\rm conv}$ (black).
       The dark purple line is a running 25th percentile of the flux over 100 frequency bins.
     }
     \label{fig:fit_flux_ell10}
\end{figure*}

If $d$ is large, the waves see a smooth convective-radiative transition and their flux depends on the profile of $N^2$ in the transition region.
\citet{Lecoanet13} consider two smooth cases:
first, an abrupt but continuous transition described by a piecewise linear profile for $N^2$, which yields the scaling relation (see their Eq.~(42))
\begin{equation}
\frac{\dif F^{\rm L}}{\dif \ln \omega \dif \ln k_{\rm h}}  \propto  k_h^{13/3} \omega^{-41/6}d^{1/3}.
\label{eq:fluxL_L13}
\end{equation}

They also consider a smoother transition given by a tanh profile for $N^2$, which yields  (see their Eq.~(41))
\begin{equation}
\frac{\dif F^{\rm T}}{\dif \ln \omega \dif \ln k_{\rm h}}  \propto  k_h^5 \omega^{-15/2}d.
\label{eq:fluxT_L13}
\end{equation}
However, the theory used to derive Eqs.~\eqref{eq:fluxD_L13}--\eqref{eq:fluxT_L13} relies on three-dimensional assumptions, for example on the counting of modes and the turbulence spectrum.
Even though we should not expect these results to hold in two-dimensional, they were still found to provide a good match to numerical spectra from two-dimensional simulations \citep{Lecoanet2021}. \\

The second main excitation mechanism of IGWs is penetrative convection.
Recently, \citet{Pincon2016} proposed a model that predicts a Gaussian energy flux spectrum that scales as (see their Eq.~(39))
\begin{equation}
\frac{\dif F^{\rm P}}{\dif \ln \omega \dif \ln k_{\rm h}}  \propto \rm e^{-\omega^2/4\nu_{\rm p}^2},
    \label{eq:flux_P16}
\end{equation}
with $\nu_{\rm p}$ the characteristic frequency associated with the plumes lifetime.
This frequency $\nu_{\rm p}$ is difficult to estimate with precision, but \citet{Pincon2016} suggest that $\nu_{\rm p} \sim \omega_{\rm conv}$ is a good first approximation.

For our simulations, we estimate the total wave energy flux $F$ using a method similar to the one of \citet{Couston2018}.
The total flux $F$ is obtained by summing the fluxes $\Fsingle$ of all individual IGW modes:
\begin{equation}
    F(r) \simeq \sum_{\omega, k_{\rm h}}  \frac{\Fsingle}{\domdkh} \,\domdkh,
\end{equation}
where $\delta \omega$ and $\delta k_h$ are the spacing of modes in spectral space, which are defined as
\begin{equation}
    \delta \omega = \frac{1}{T_{\rm s}} \quad;\quad \delta k_h = \frac{1}{r}
    \label{eq:mode_spacing}
,\end{equation}
where $T_s$ is the sampling time.
Then replacing $\Fsingle$ by its expression given by Eq. \eqref{eq:wave_flux}, we have
\begin{equation}
    F(r) \simeq \sum_{\omega, k_{\rm h}}  \frac{\frac{1}{2}\rho \frac{N}{k_{\rm h}} P[\hat{\vel}_r^2](r,\omega,\ell)}{\domdkh} \,\domdkh.
    \label{eq:flux_tot_spectral}
\end{equation}
We can now introduce the discrete version of the wave energy flux:
\begin{equation}
\sum_{\omega, k_{\rm h}}  \frac{\frac{1}{2}\rho \frac{N}{k_{\rm h}} P[\hat{\vel}_r^2](r,\omega,\ell)}{\domdkh} \,\domdkh \coloneqq
\sum_{\omega, k_{\rm h}} \frac{\delta F}{\domdkh} \,\domdkh.
\label{eq:discrete-flux}
\end{equation}
This last form can be related to a continuous form of the differential wave energy flux:
\begin{equation}
\sum_{\omega, k_{\rm h}} \frac{\delta F}{\domdkh} \,\domdkh \simeq
\int \frac{\dif F}{\dif \omega \dif k_{\rm h}} \dif \omega \dif k_{\rm h}.
\label{eq:discrete-cont-flux}
\end{equation}
In order to compare our estimation to the analytical expressions introduced above,
we identify from Eqs. \eqref{eq:discrete-flux} and \eqref{eq:discrete-cont-flux}
\begin{equation}
\frac{\dif F}{\dif \omega \dif k_{\rm h}}
= \frac{1}{2}\rho \frac{N}{k_{\rm h}} \frac{P[\hat{\vel}_r^2]}{\domdkh}.
\end{equation}
Then, using the expressions introduced in Eq.~\eqref{eq:mode_spacing} we finally obtain
\begin{equation}
\frac{\dif F}{\dif \ln \omega \dif \ln k_{\rm h}}
=
\omega k_{\rm h}
\frac{\dif F}{\dif \omega \dif k_{\rm h}}
\sim
\frac{1}{2}\rho T_{\rm s} r N  \omega P[\hat{\vel}_r^2].
\label{eq:flux_MUSIC}
\end{equation}

The fluxes are calculated at a radius far away from the convective boundary ($r \sim r_{\rm conv} - l_{\rm max}$) to ensure that $\hat{\vel}_r$ only captures the wave motions and not additional motions due to convective penetration.
These fluxes extracted from the numerical data can then be compared to the theoretical spectra predicted by \citet{Lecoanet13} and \citet{Pincon2016}.
The results are shown in Fig.~\ref{fig:fit_flux_ell10} for $\ell = 10$. We recall that the relation between $k_{\rm h}$ and $\ell$ is defined by Eq. \eqref{eq:kh}.

The four spectra present similar shapes, with a flat part at low frequencies and a peak that is shifted towards higher frequencies with increasing luminosity enhancement factor.
The peak follows the convective turnover frequency, which is indicated by the vertical grey line.
For a better visual comparison with theoretical power laws, we also compute and plot a running median on the 25th percentile of the flux with a window of 100 frequency bins (dark purple line).
According to \citet{Lecoanet13}, the scaling relationships Eqs.~\eqref{eq:fluxD_L13}--\eqref{eq:fluxT_L13} are only valid for frequencies $\omega \geq \omega_{\rm conv}$.
In the high frequency part, the baseline of the spectra approximately follows a scaling between $\omega^{-13/2}$ and  $\omega^{-15/2}$, broadly consistent with the theoretical predictions;
however, determining the precise slope is difficult because of the presence of strong g modes,
which appear as `combs' of high-energy peaks.
We find similar behaviours for other values of $\ell$.
Following the arguments of \citet{Lecoanet13} and given the increase in the overshooting length characterising the convective-radiative transition with the luminosity enhancement factor,
one would expect the high-$\omega$ slope to evolve as luminosity increases,
from the discontinuous case of Eq.~\eqref{eq:fluxD_L13} to the smoother case of Eq.~\eqref{eq:fluxT_L13}.
However, given that the theoretical slopes are quite close to each other, it is difficult to determine the exact slope of the calculated spectra beyond $\omega_{\rm conv}$ from such a comparison.

In Fig.~\ref{fig:fit_flux_ell10}, we have also plotted part of the theoretical spectrum predicted by \citet{Pincon2016}, taking $\nu_{\rm p} = \omega_{\rm conv}$.
To observe the effect of varying this characteristic frequency, we have also plotted a spectrum with $\nu_{\rm p} = \sqrt{2}\omega_{\rm conv}$;
as expected, increasing $\nu_{\rm p}$ increases the width of the spectrum.
The spectrum of \citet{Pincon2016} is consistent with the shape of our simulated spectra in the vicinity of $\omega_{\rm conv}$.
The agreement seems to improve as the luminosity enhancement factor is increased.
This result suggests that both Reynolds stresses and plume penetration play a role in IGW generation.

We note that the larger the luminosity enhancement factor of the stellar model, the closer the peak of the spectrum lies to the Brunt-Väisälä frequency.
The luminosity enhancement thus limits the range of frequencies over which numerical models can be compared to theoretical predictions.
As noted in Paper I, very large enhancement factors for the luminosity, with factors  $> 10^6$, will produce unrealistic results because of convective velocities in the outer part of the domain that could become close to the speed of sound.
In addition, this paper highlights that for such large factors the convective turnover frequencies, which scale as $L^{1/3}$, would become higher than the Brunt-Väisälä frequency.

\begin{figure}[h!]
  \centering
  \includegraphics[width=9cm]{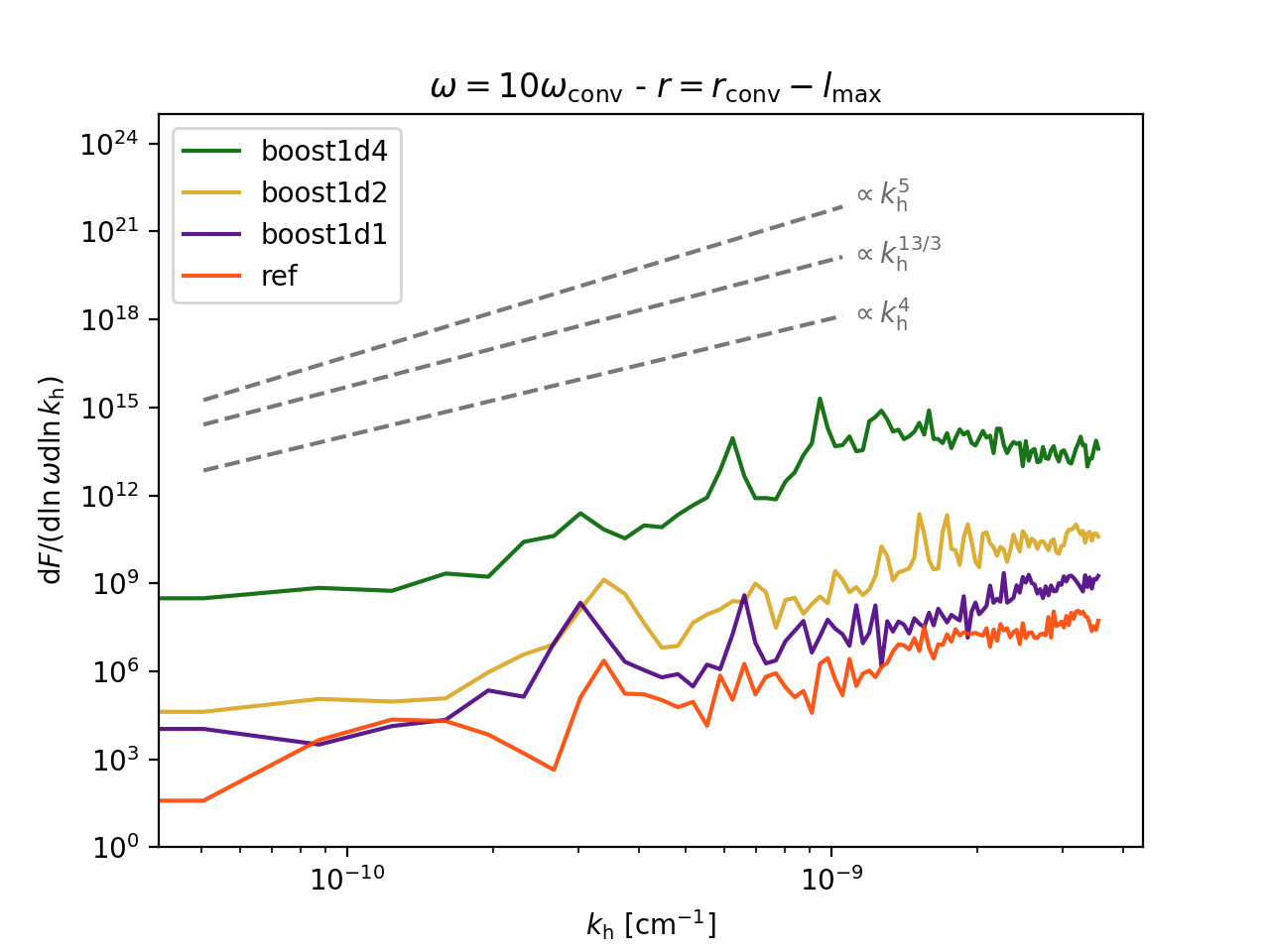}
  \caption{
    Wave energy flux as a function of the horizontal wavenumber for the four simulations at depth $r =r_{\rm conv}-l_{\rm max}$.
    The frequency is fixed at $\omega = 10\omega_{\rm conv}$.
    The dashed lines represent the three theoretical scaling laws, Eqs.~\eqref{eq:fluxD_L13}--\eqref{eq:fluxT_L13}, from \citet{Lecoanet13}.
  }
  \label{fig:fit_flux_freq40}
\end{figure}

We also analyse the horizontal wavenumber dependence of the wave flux derived from the simulations.
In Fig.~\ref{fig:fit_flux_freq40}, the wave energy flux of Eq.~\eqref{eq:flux_MUSIC}
is displayed for the four simulations as a function of $k_{\rm h}$ at a frequency $\omega = 10 \omega_{\rm conv}$ for the relevant simulation.
This allows a meaningful comparison between the different simulations.
Figure \ref{fig:fit_flux_freq40} shows that the four numerical models present similar features.

The scaling relationships Eqs.~\eqref{eq:fluxD_L13}--\eqref{eq:fluxT_L13} from \citet{Lecoanet13} are also displayed in Fig.~\ref{fig:fit_flux_freq40}.
The derivation of their expressions is valid for $k_{\rm h} \lesssim k_{\rm h,max} \propto \omega_{\rm conv}^{-3/2}$ (see Eq.~\eqref{eq:ell_max}).
The low-$\omega$ slope of the simulated spectra is roughly consistent with these theoretical scalings,
even though the large statistical variance of the spectra prevents an accurate comparison without running over very long simulation times,
or averaging over a large number of ensemble simulations.

In summary, the wave fluxes in the radiative zone obtained from present simulations obey scaling laws with frequency and wavenumber that are broadly consistent with the ones predicted by theoretical models.
However, discriminating or constraining the theoretical models with precise measurement of slopes from simulations remains challenging.
In particular, even though our simulations are consistent with theoretical predictions, we caution that these simulations are not fully turbulent and that the two-dimensional geometry does not allow us to resolve realistic plumes (see Sect. \ref{discussion} for further discussion).

\subsection{Spatial damping of IGWs}
\label{sec:spatial_damping}

\begin{figure*}
\centering
   \includegraphics[width=0.49\textwidth]{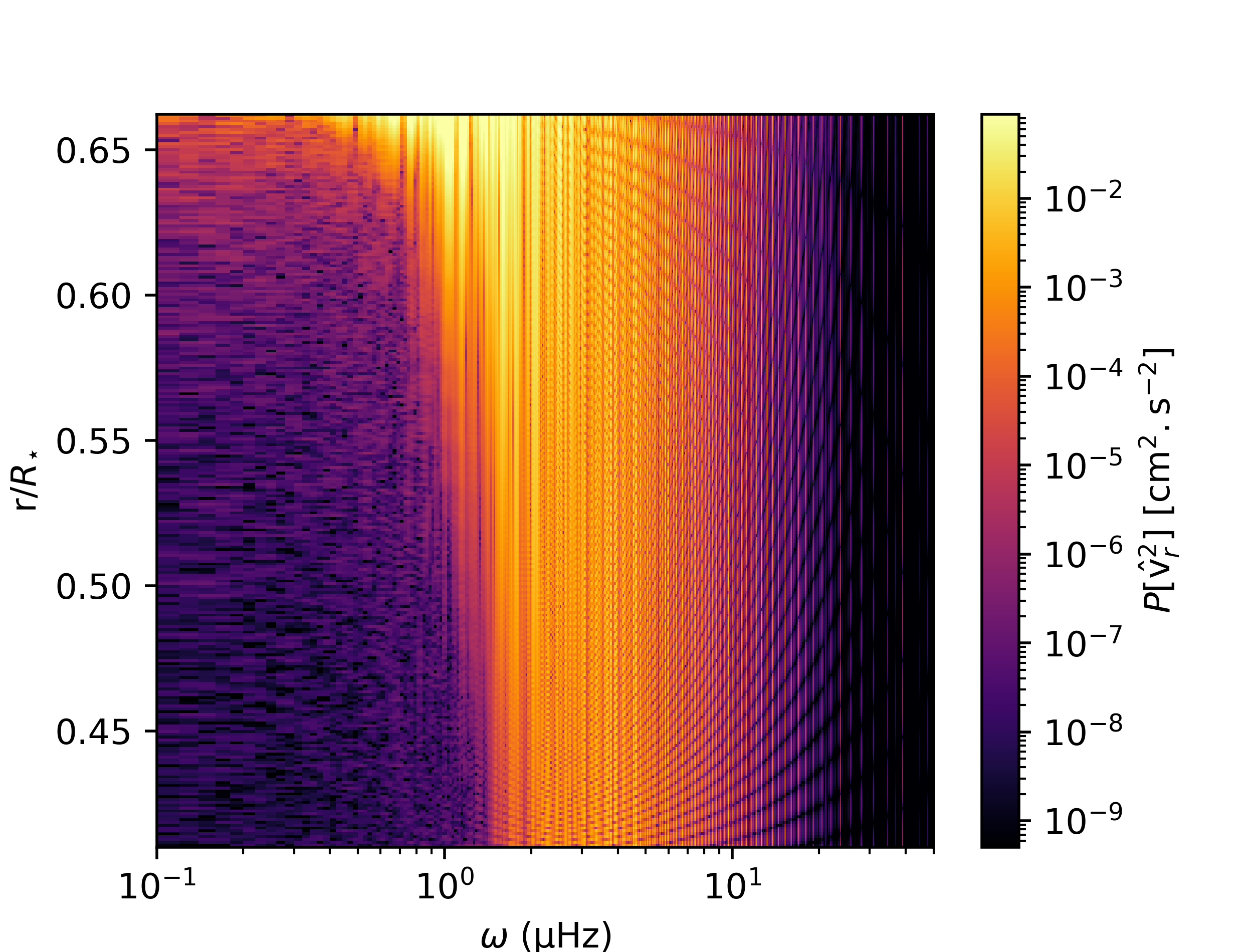}
   \includegraphics[width=0.49\textwidth]{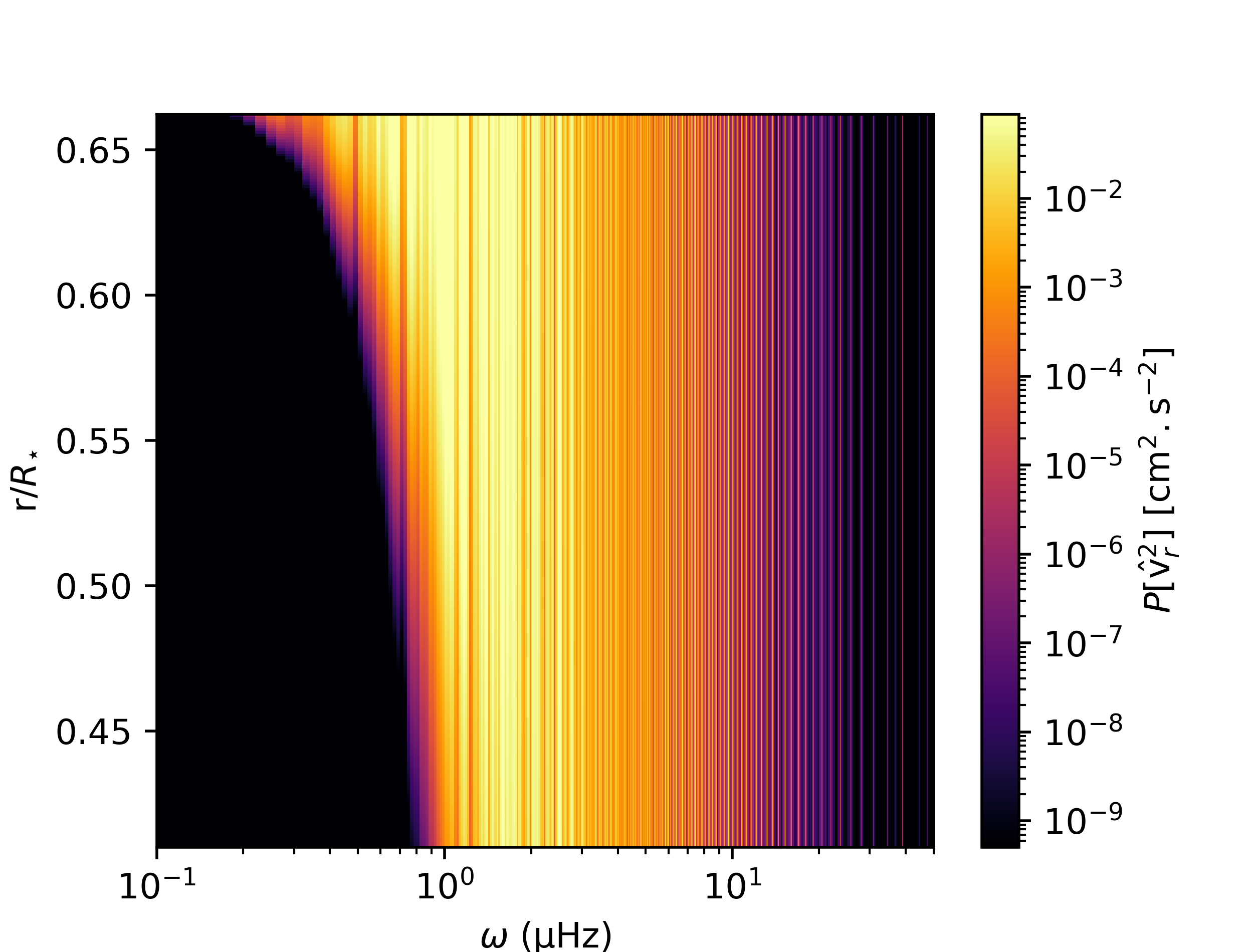}
   \includegraphics[width=0.49\textwidth]{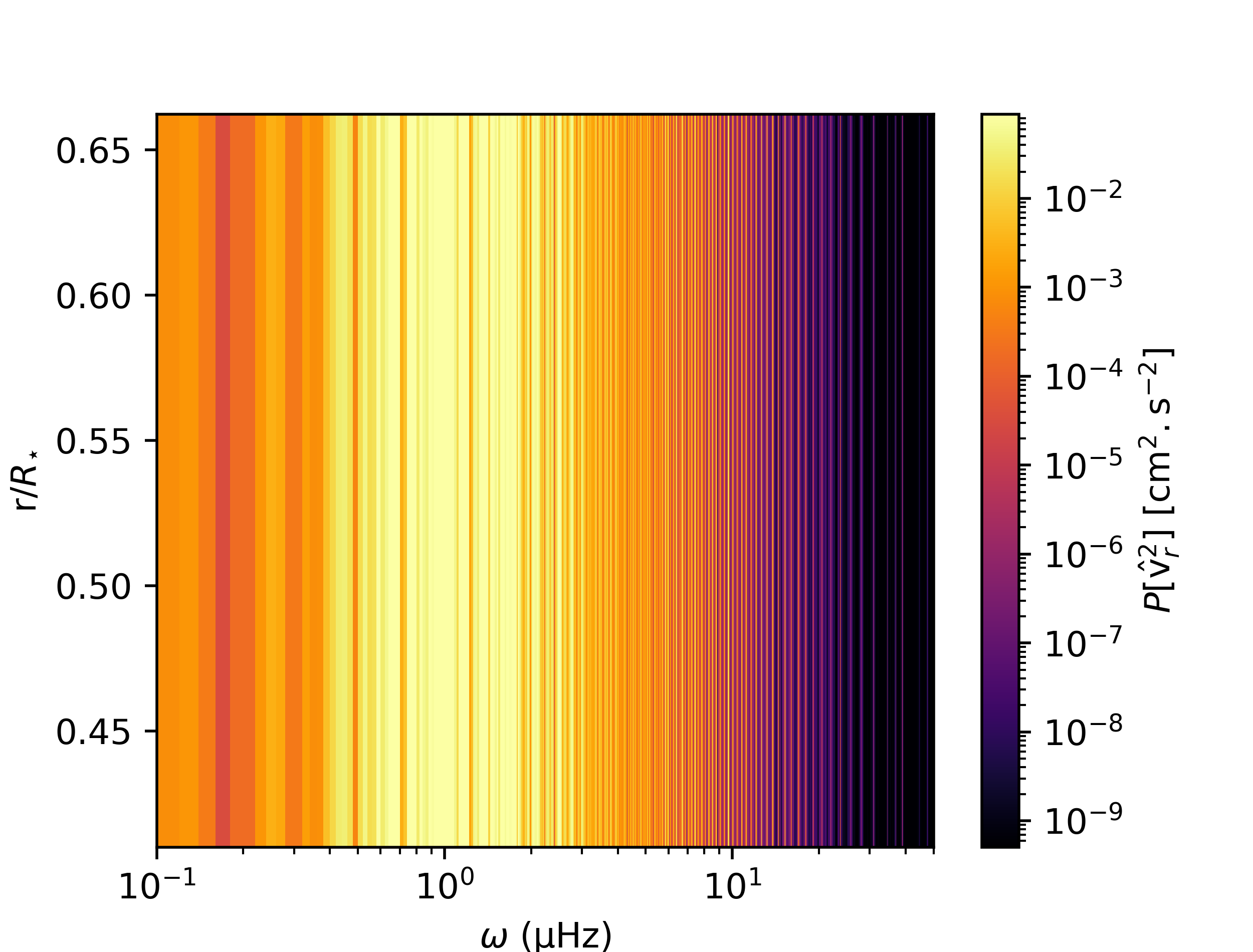}
   \includegraphics[width=0.49\textwidth]{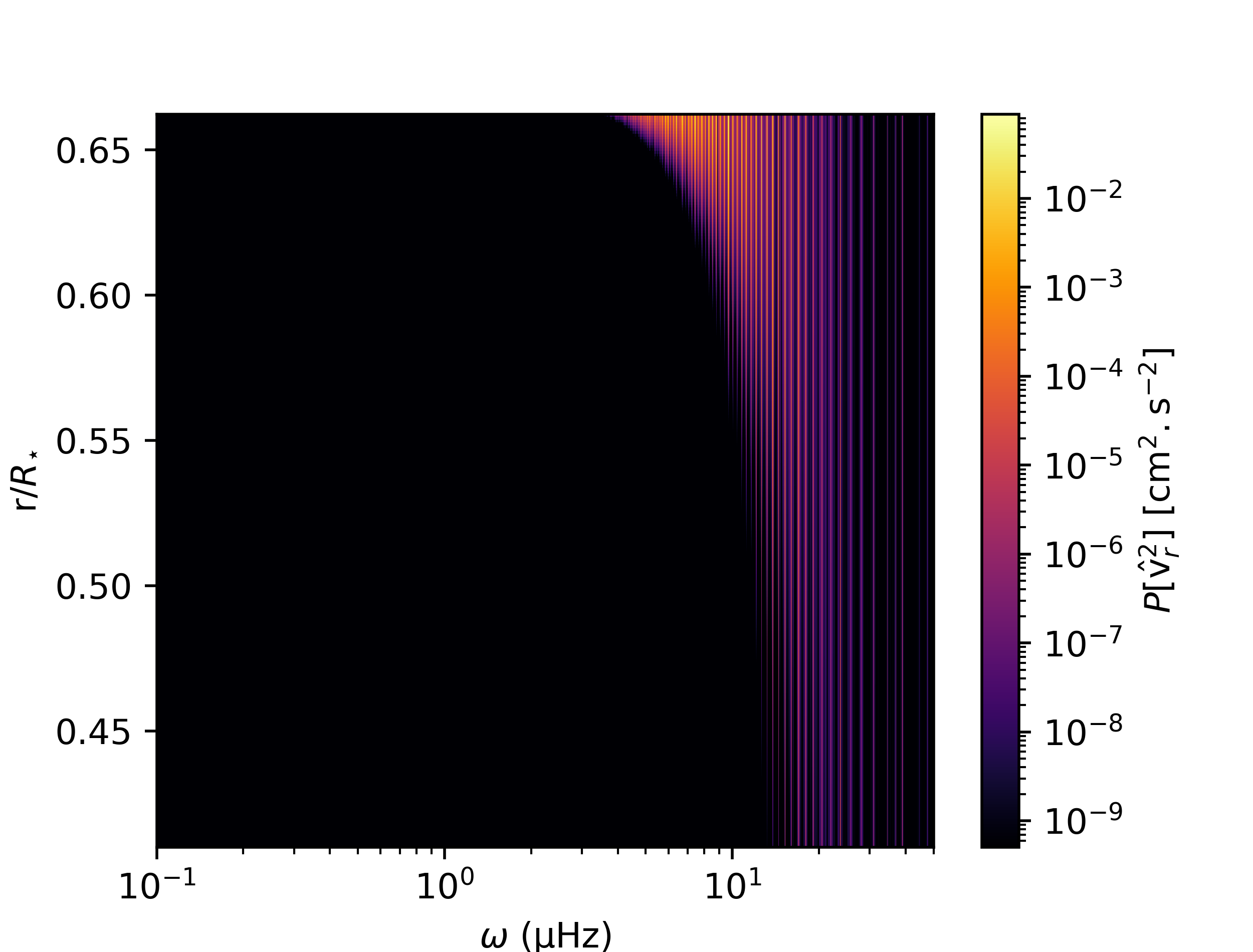}
     \caption{Power spectra of the radial velocity for the simulation \textit{ref} obtained from MUSIC (top left) and the theoretical prediction using Eq. \eqref{eq:radiative_damping} with the dependence on frequency, $\omega^{-n}$, which varies: $n=4$ (top right), $n=3$ (bottom left), and $n=5$ (bottom right). The angular degree is fixed at $\ell$ = 5. For the theoretical spectra, a minimal threshold is set at a value of $10^{-7}$ for better visibility.}
     \label{fig:damping_krad}
\end{figure*}

Internal gravity waves propagating in stably stratified stellar interiors are expected to be damped by radiative diffusion.
Their amplitude exponentially decays with a factor $\exp(-\tau)$, with $\tau$ given in the limit $\omega \ll N$ by \citep{Press1981, Zahn1997}
\begin{equation}
    \tau(r,\ell,\omega) = [\ell(\ell+1)]^{3/2} \int^{r_{\rm e}}_r \kappa_{\rm rad} \frac{N^3}{\omega^4} \frac{\dif r}{r^3},
    \label{eq:radiative_damping}
\end{equation}
where $r_e$ is the radius at which the waves are excited. In Eq. \eqref{eq:radiative_damping}  $\kappa_{\rm rad} = \chi/(\rho c_p)$ is the radiative diffusivity with $\chi$ the radiative conductivity defined by Eq.~\eqref{eq:radiative_flux}.

In order to verify that the damping of the waves in the simulations is consistent with the theoretical expectations, in Fig.~\ref{fig:damping_krad} we use a similar method as \citet{Alvan2014} to compare the evolution of power spectrum of the radial velocity, $P[\hat{\vel}_r^2]$ (see Eq. \eqref{eq:def_power_spectrum}), to the one predicted theoretically, $P_{\rm theory}[\hat{\vel}_r^2]$, for a fixed harmonic degree $\ell_0 = 5$,
which is defined as
\begin{equation}
  P_{\rm theory}[\hat{\vel}_r^2](r,\ell_0,\omega) = P[\hat{\vel}_r^2](r_{\rm conv} - l_{\rm max},\ell_0 ,\omega) \times  {\rm e}^{-\tau(r,\ell_0,\omega)}.
  \label{eq:damping_p}
\end{equation}
The depth $r = r_{\rm conv} - l_{\rm max}$ is chosen in order to avoid any contribution to the radial velocity from convective penetration.
In the integral \eqref{eq:radiative_damping}, the upper bound $r_{\rm e}$ is thus set to $r_{\rm conv} - l_{\rm max}$.

To visually compare the spectral density plots, we set an identical colour scale for all panels of Fig.~\ref{fig:damping_krad}, adjusted to the noise floor of the simulation.
The simulation (top left panel) and theoretical $\omega^{-4}$ scaling (top right panel) spectra in Fig.~\ref{fig:damping_krad} show similar patterns,
with g modes (vertical bright lines) formed in the radiative zone for frequencies $\gtrsim \uHz{1}$,
as well as the absence of modes below $\sim \uHz{0.5}$.
We recall that in order to form a g mode, waves have to travel to an inner turning point, where they are reflected and then travel back outwards to an outer turning point.
The reflections and interference of IGWs between these two turning points will form g modes \citep{Alvan2015}.

The radial profile of the spectra between $\sim \uHz{0.5}$ and $\sim \uHz{10}$ are also similar in both top panels of Fig. \ref{fig:damping_krad}, with waves propagating deeper as the frequency increases;
in addition, both spectra reach the noise floor (darkest colour) at comparable frequencies for each given radius in both top panels.
As expected from the radiative damping formulation, the damping is stronger for low frequency waves.
In the MUSIC spectrum, the nodes of the g modes are visible as dark spots of low energy spaced in radius along a given mode; they are not visible on the theoretical reconstructions, as expected from our definition of $P_{\rm theory}[\hat{\vel}_r^2]$.

\begin{figure*}
\centering
   \includegraphics[width=0.33\textwidth]{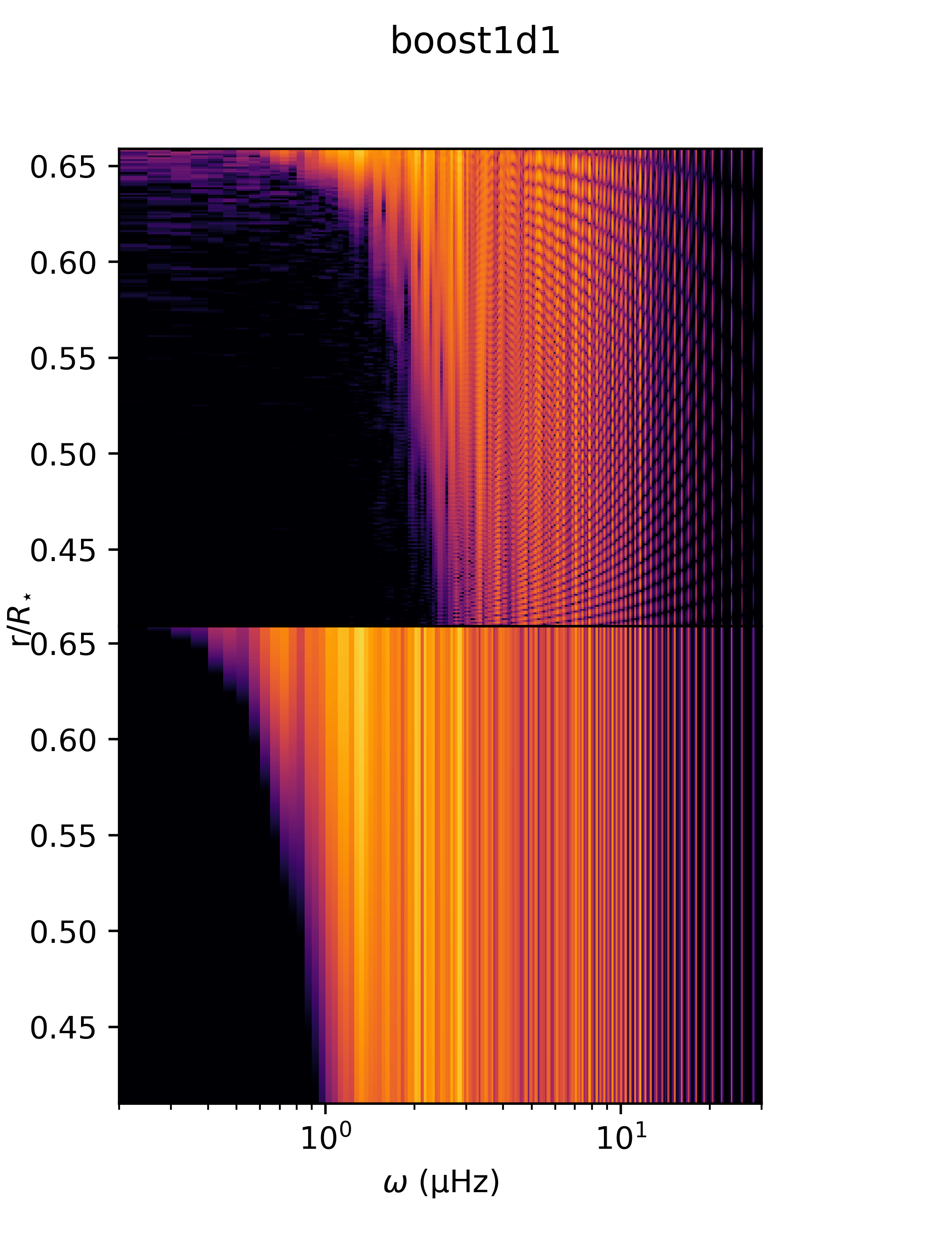}
   \includegraphics[width=0.33\textwidth]{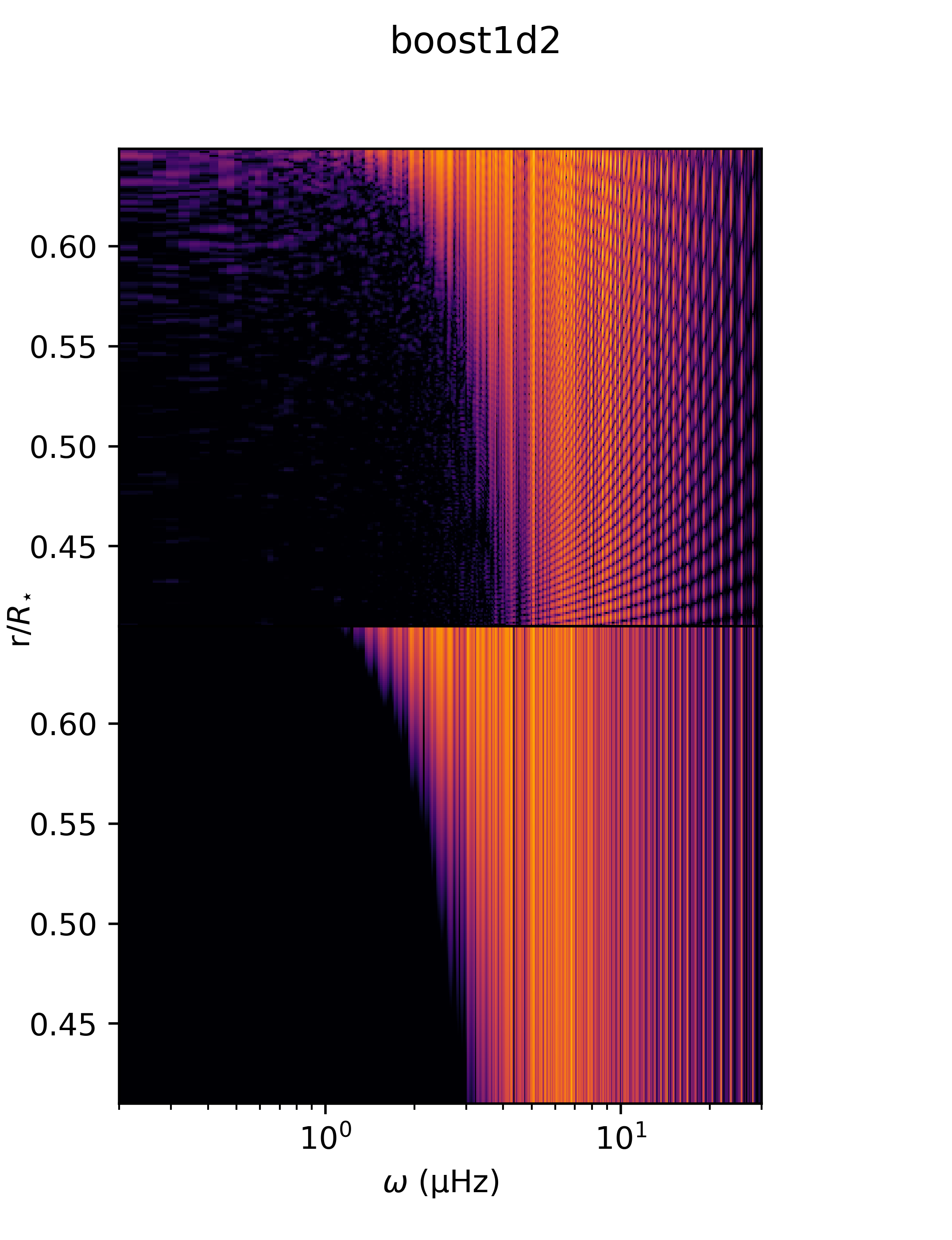}
   \includegraphics[width=0.33\textwidth]{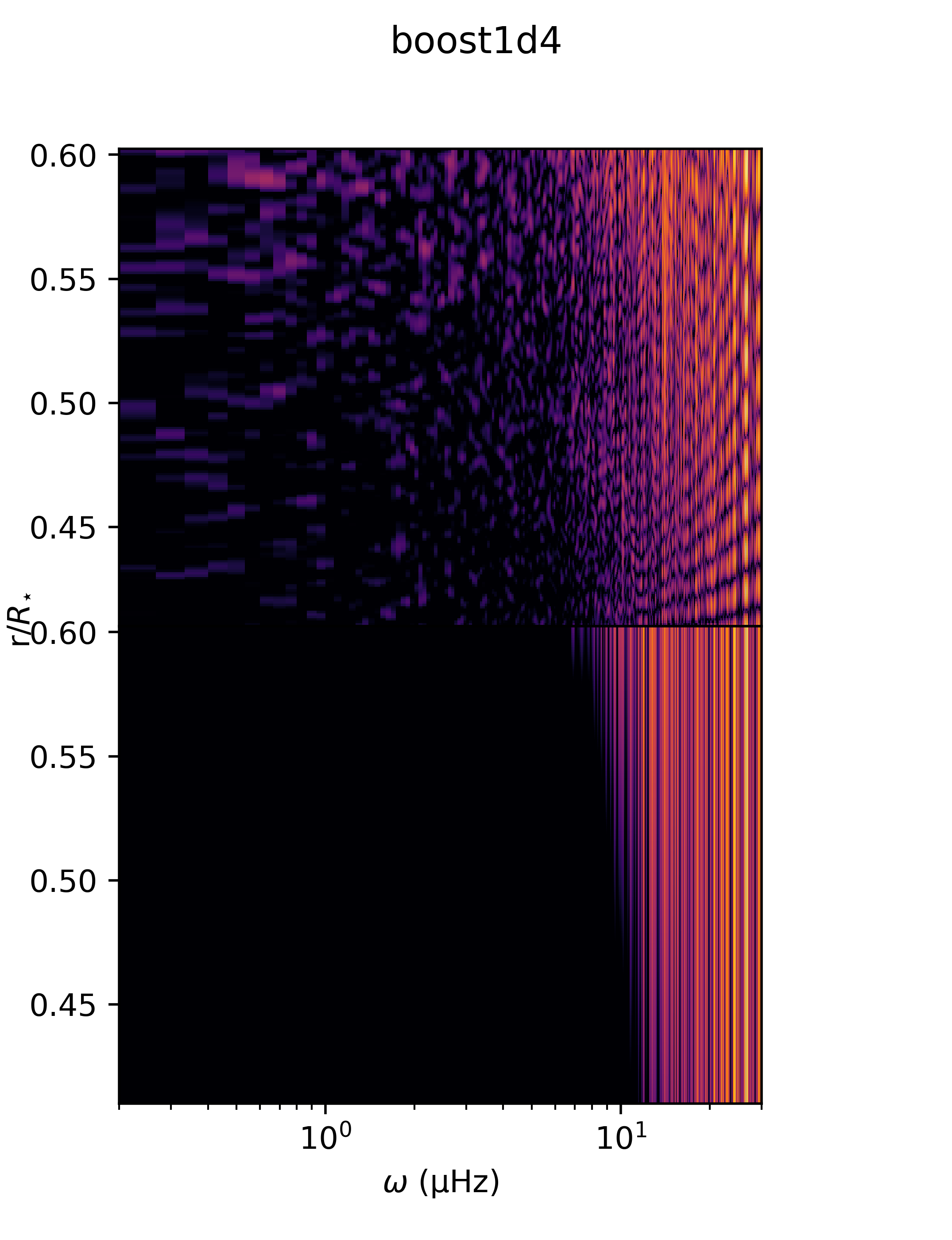}
     \caption{Power spectra of the radial velocity for simulations \textit{boost1d1} (left), \textit{boost1d2} (middle), and \textit{boost1d4} (right) obtained from MUSIC (top row) and the theoretical prediction using Eq. \eqref{eq:radiative_damping} with the dependence on frequency, $\omega^{-4}$ (bottom row). The angular degree is fixed at $\ell$ = 5. For the theoretical spectra, a minimal threshold is set for better visibility.}
     \label{fig:damping_boost}
\end{figure*}

In previous studies of IGWs using hydrodynamical simulations, the radiative damping of the waves was found to have a scaling with frequency closer to $\omega^{-3}$, instead of $\omega^{-4}$ \citep{Rogers2013, Alvan2014}.
In order to directly compare our results with \citet{Rogers2013, Alvan2014} and to keep $\tau$ dimensionless when changing only the exponent of $\omega$, we replace $N^3 \omega^{-4}$ in \eqref{eq:radiative_damping} with $N^3 \omega^{-q} \times \left( 1~ \rm{Hz} \right)^{q-4}$ for $q=3,4,5$.
We plot in Fig. \ref{fig:damping_krad} the theoretical spectra calculated using a dependence of $\omega^{-3}$ and $\omega^{-5}$. We can exclude these scaling laws for the MUSIC spectrum, which has a dependence in  $\omega$ closer to $\omega^{-4}$ as predicted by Eq. \eqref{eq:radiative_damping} (see Sect. \ref{discussion} for further discussion).\\

Figure \ref{fig:damping_boost} shows that Eq.~\eqref{eq:radiative_damping} also holds for the three boosted simulations,
provided that radiative diffusivity in Eq.~\eqref{eq:radiative_damping} is increased with the same enhancement factor as the luminosity.
In addition to simulation \textit{ref}, we also verified for these boosted simulations that damping laws scaling as $\omega^{-3}$ and $\omega^{-5}$ can be excluded.
The MUSIC spectra thus provide a good agreement with the theoretical predictions of radiative damping.
We further discuss this point in Sect. \ref{discussion}.
%%%

These results show that if both the luminosity and the radiative diffusivity are enhanced by the same factor, the damping of the waves follows similar scaling for all enhancement factors, and is in agreement with the theoretical prediction for radiative damping.
The boosting factor will have an important effect on the spectrum of surviving g modes. Since g modes can only appear with waves propagating in the whole domain, they can only be observed when waves are able to travel down to the inner boundary without being fully damped.
Because increasing the luminosity enhancement factor shifts the IGW excitation spectrum towards higher frequencies,
the frequency of waves surviving at the inner boundary will increase with the boosting factor.
Indeed, for \textit{ref} in Fig.~\ref{fig:damping_krad}, the lowest frequency g modes that are formed have a frequency of $\omega \sim \uHz{1.0}$.
In Fig.~\ref{fig:damping_boost}, we can see that for \textit{boost1d1}, \textit{boost1d2,} and \textit{boost1d4},
the minimum frequency of the surviving g modes is $\omega \sim \uHz{2.0}$, $\omega \sim \uHz{3.0},$ and $\omega \sim \uHz{8.0}$, respectively, for the harmonic degree $\ell = 5$.
Below these frequencies, waves are damped before being able to propagate back and forth in the radiative zone, and thus are unable to form g modes. \\

\begin{figure}
\centering
   \includegraphics[width=0.4\textwidth]{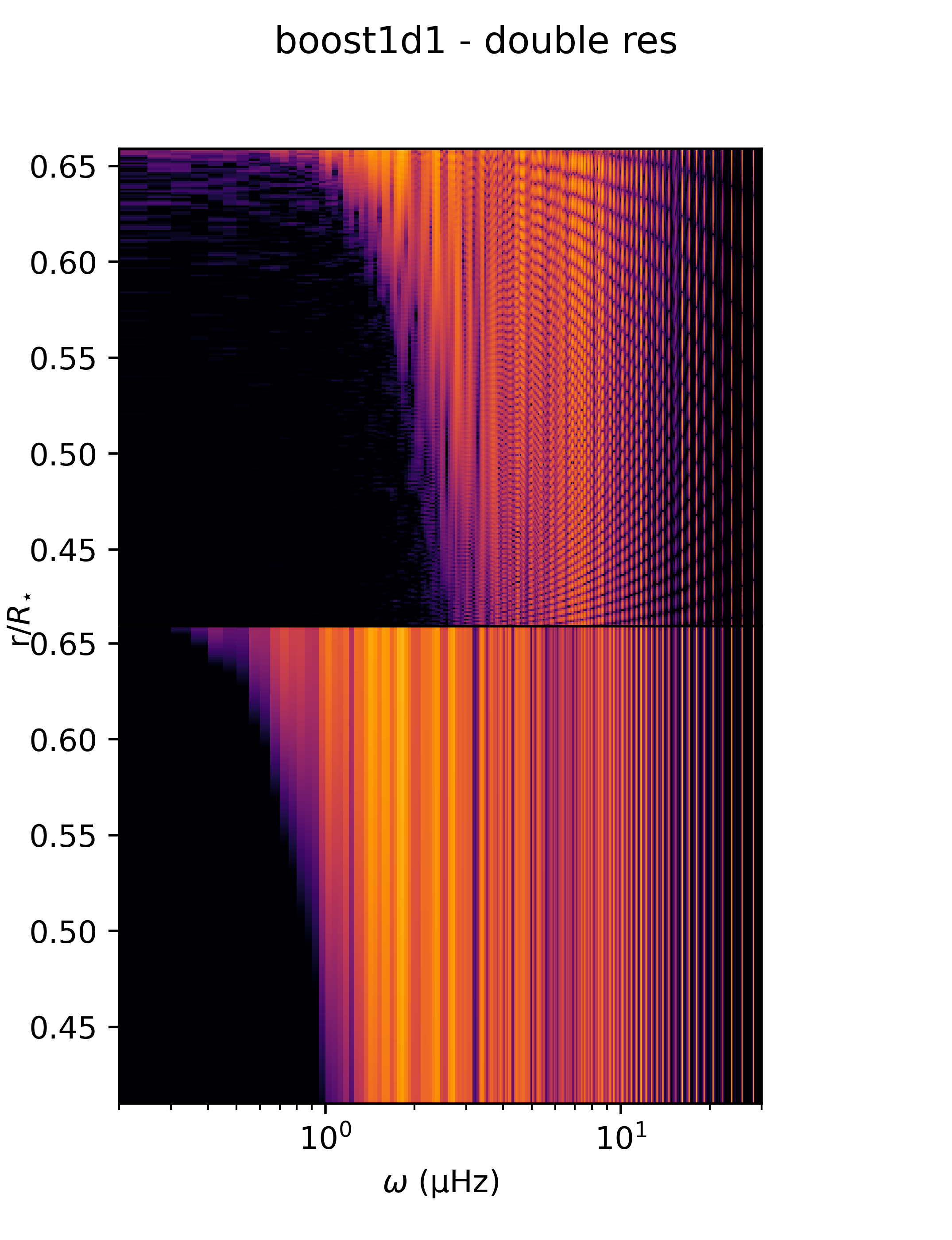}
     \caption{Power spectra of the radial velocity for the simulation \textit{boost1d1} with double resolution (1024x1024) obtained from MUSIC (top) and the theoretical prediction using Eq. \eqref{eq:radiative_damping} with the dependence on frequency, $\omega^{-4}$ (bottom). The angular degree is fixed at $\ell$ = 5. For the theoretical spectra, a minimal threshold is set for better visibility.}
     \label{fig:damping_HRes}
\end{figure}

Finally, we would like to draw attention to a numerical resolution issue related to the study of IGW damping in simulations.
The condition of propagation of IGWs in a stably stratified medium is their dispersion relation given by Eq.~\eqref{eq:dispersion},
which predicts the radial wavelength to become small as $\omega \to 0$.
For a given spatial grid, there is a maximal radial wavenumber, or equivalently a minimal wavelength, that can be resolved on the grid.
As the wavelength approaches the grid resolution, numerical dissipation of the numerical scheme is expected to increase;
ultimately, as the radial wavelength approaches twice the length of a grid cell, it becomes impossible to represent the wave on the grid, and aliasing occurs.
This is potentially problematic for the study of IGW damping, which requires looking at very low frequencies.

For our simulations, we can see in Figs.~\ref{fig:damping_krad} and \ref{fig:damping_boost} that some aliasing is present in the g modes around 1 or $\uHz{2}$.
In order to confirm our results, we used the test simulation for \textit{boost1d1} with double resolution (1024x1024) already introduced in paper I.
Figure \ref{fig:damping_HRes} shows the power spectrum from MUSIC together with the theoretically predicted one.
There is no more aliasing at low frequencies, and more importantly, the spectrum is still very similar to the one shown in Fig.~\ref{fig:damping_boost} for simulation \textit{boost1d1} with lower resolution.
This gives us thus some confidence that the damping we observe in our simulations is indeed radiative damping.

%%%%%%%%%%%%%%%%%%%%%
%   SECTION 7
%%%%%%%%%%%%%%%%%%%%%

\section{Discussion and conclusion}
\label{discussion}

Following Paper I, this study focuses on the impact of enhanced luminosity and thermal diffusivity on IGWs excited by convection in hydrodynamical simulations of stellar interiors.
The results confirm the suggestion of \citet{Lecoanet2019}
that artificially increasing the luminosity and hence the wave amplitudes can impact the wave properties.
Increasing the luminosity of a simulation decreases the convective turnover time, $\tau_{\rm conv}$, and thus increases the convective turnover frequency, $\omega_{\rm conv}$.
The larger the luminosity, the higher $\omega_{\rm conv}$.
Typical dynamical quantities such as velocities scale with the luminosity in the convective zone.
We show that an appropriate rescaling of the frequencies with $\omega_{\rm conv}$ and of the velocities with $L^{1/3}$ provides a kinetic energy spectrum in the convective zone that is independent of the luminosity enhancement factor.
But firstly, such rescaling does not hold in the radiative zone.
Secondly, our results highlight that the relevant frequency range of IGWs that are excited and propagating strongly depends on the luminosity enhancement factor.
The waves that bear most of the energy are not in the same frequency and spatial ranges for simulations with different luminosities.
As the luminosity increases, the frequency distribution is shifted towards higher $\omega$, whereas the angular degree distribution is shifted towards lower $\ell$ (larger wavelengths).

Our simulated energy flux spectra are broadly consistent with IGWs being generated from a combination of two excitation mechanisms: Reynolds stresses and penetrative convection.
At high $\omega$, we observe scaling laws for the flux that are compatible with the Reynolds stress excitation models from \citet{Lecoanet13}.
While those authors predict different expressions for the wave energy flux depending on the stratification at the radiative--convective boundary,
it is difficult to compare the slopes obtained from the simulations with the ones predicted by theoretical models
given the variance of the spectra and the presence of high-amplitude g modes.
As $\omega$ decreases, the radial energy flux departs from a power law and reaches a maximum for $\omega \simeq \omega_{\rm conv}$. In this range the flux is more consistent with the predictions from \citet{Pincon2016} from penetrative convection.

Care must be taken, however, in interpreting IGW excitation processes from two-dimensional simulations:
with global two-dimensional simulations of stellar interiors, flows cannot present a real state of turbulence, feature well-identified plumes, or finely resolve penetrating convection.
Therefore, we cannot expect the analytical models of \citet{Lecoanet13} and \citet{Pincon2016} to apply precisely and quantitatively.
Nevertheless, our results qualitatively support a picture where IGWs are simultaneously excited by Reynolds-like stresses at high $\omega$
and by penetrating convection around $\omega_{\rm conv}$.
Our work therefore suggests an important consequence for boosted luminosity simulations:
as the luminosity enhancement factor increases, the wave flux is shifted towards higher $\omega$.
In addition, the peak of the wave flux will be closer to the Brunt-Väisälä frequency, limiting the extent of the frequency range of excited IGWs.
Furthermore, as mentioned in Paper I and Sect. \ref{radial_flux}, larger enhancement factors could impact the local stratification,
further interfering with wave excitation processes.
Consequently, while it is difficult with present simulations to conclude on the impact of the luminosity enhancement factor on the detailed {shape} of the wave flux spectrum,
our results highlight that the artificial boosting of a simulation has a noticeable impact on the wave flux by changing the {location} of the peak of the spectra in terms of frequency and horizontal wavenumber.

For radiative damping at low frequencies, in contrast to results reported by other groups, our simulations present a decay in the amplitude of waves in the radiative zone, which closely decreases as $\propto \exp(-\omega^{-4})$ as predicted by theory \citep{Press1981}. This dependence of the damping on frequency holds in the boosted simulations on the condition that the radiative diffusivity is increased by the same amount as the luminosity.
We also show that the waves that reach the inner boundary of the domain are affected by the boost
as a result of a change in the wave excitation spectrum and with fewer g modes surviving with higher luminosity enhancement factors.
This could have some importance when studying angular momentum transport in stellar interiors.
We want to stress that radiative damping is difficult to study numerically because it is more efficient at low frequencies,
at which the radial wavelength of the IGWs becomes close to the radial grid resolution due to the dispersion relation.
Our test simulation at double resolution strengthens our confidence that we are actually observing radiative damping.

In summary, our analysis shows that artificially increasing the luminosity of a stellar model can be a useful technique.
It can be justified in particular for studies restricted to the dynamics in a convective zone since appropriate rescaling laws can apply.
However, it has to be used with great caution to predict the spectra of IGWs and g modes in stars. These spectra will define how waves interact with the dynamics and internal structure of stars. Consequently, making predictions for more realistic systems, in particular when related to mixing and angular momentum transport, is not straightforward when using simulations with artificially enhanced luminosity.

 \section{Acknowledgements}
This work is partly supported by the consolidated STFC grant ST/R000395/1 and the ERC grant No. 787361-COBOM. The authors would like to acknowledge the use of the University of Exeter High-Performance Computing (HPC) facility ISCA and of the DiRAC Data Intensive service at Leicester, operated by the University of Leicester IT Services, which forms part of the STFC DiRAC HPC Facility. The equipment was funded by BEIS capital funding via STFC capital grants ST/K000373/1 and ST/R002363/1 and STFC DiRAC Operations grant ST/R001014/1. DiRAC is part of the National e-Infrastructure. A.S.B. acknowledge funding by ERC WHOLESUN 810218 grant.

\bibliographystyle{aa}
\bibliography{references}

%\clearpage
\begin{appendix}
\section{Spherical harmonics and Fourier amplitudes}
\label{apdx:sh-ft}

In two-dimensional spherical geometry, on the unit axisymmetric sphere $\mathcal{S}$ parameterised by the polar angle $\theta$,
we define the spherical harmonics coefficients $\hat f_\ell$ of a function $f(\theta)$ from the expansion
\begin{align}
  f(\theta) &= \sum_{\ell \geq 0} \hat f_\ell \; Y_\ell^0(\theta), \\
  Y_\ell^0(\theta) &= \sqrt{\frac{2\ell+1}{4\pi}} P_\ell(\cos \theta),
\end{align}
where $P_\ell$ is the $\ell$-th Legendre polynomial.
This expansion is identical to the usual spherical harmonics on the sphere, but restricted to the axisymmetric mode $m=0$.
The choice of normalisation makes the $Y_\ell^0$ an orthonormal basis for the $L^2$ inner product on $\mathcal{S}$
\begin{align}
  \int_\mathcal{S} Y_\ell^{0*} \, Y_{\ell'}^0 \; \mathrm{d}\Omega
  = 2\pi \int_0^\pi Y_\ell^0 \, Y_{\ell'}^0 \sin \theta \; \mathrm{d}\theta
  = \delta_{\ell \ell'}.
\end{align}

The Fourier coefficients $\tilde f_k$ of a temporal signal $f(t)$ over a period $T$ are defined from the expansion
\begin{align}
  f(t) W(t/T) &= \sum_{k} \tilde f_k \; e^{2i\pi k t/T}.
\end{align}
where $W$ is the Blackman window function, used for apodisation of the spectra of non-periodic signals $f$.
In the above formula, mode $k$ corresponds to a physical frequency $\omega_k = 2\pi |k|/T$.

The corresponding Fourier power spectrum $P[f]_n$ is defined from the $\tilde f_k$ for all non-negative frequency mode integers $n$ as
\begin{align}
  P[f]_n = \sum_{|k|=n} |\tilde f_k|^2 =
  \begin{cases}
    |\tilde f_0|^2 & \text{if $n=0$,}\\
    |\tilde f_n|^2 + |\tilde f_{-n}|^2 & \text{if $n>0$.}
  \end{cases}
  \label{eq:def_power_spectrum}
\end{align}
We note that for a real-valued signal $f$, $\forall k, |\tilde f_k|^2 = |\tilde f_{-k}|^2$.

\end{appendix}

\end{document}